\documentclass[aps,preprint,groupedaddress,floatfix]{revtex4-1}
\usepackage{graphicx}
\usepackage{natbib}
\usepackage{xcolor}

\begin{document}

\title{3D integration and packaging for solid-state qubits}
\author{D. Rosenberg, S. Weber, D. Conway, D. Yost, J. Mallek, G. Calusine, R. Das, D. Kim, M. Schwartz, W. Woods, J. L. Yoder, W. D. Oliver}
\affiliation{MIT Lincoln Laboratory}
\maketitle

\section{Quantum Processing with Solid-state Qubits}
Quantum processing has the potential to transform the computing landscape by enabling efficient solutions to problems that are intractable on classical processors. The field was sparked by a suggestion by Richard Feynman in 1981 that a controllable quantum system can be used to simulate other quantum systems, such as materials and chemistry. In the 1990's, interest in quantum computing grew rapidly with the introduction of the first quantum ``killer app''-- the potential of a large-scale quantum processor to break certain types of public encryption schemes \cite{Shor_1995}. Recently, there has been growing consensus that myriad other fields besides data security could be impacted by the development of a quantum processor, including machine learning \cite{Biamonte_2017}, many optimization problems \cite{QAOA}, and Feynman's original idea of simulation of materials properties \cite{qsim}.

The fundamental logic element  of a quantum computer is a quantum bit--termed ``qubit." Unlike a classical bit, which can take on one of two possible values, ``0'' or ``1'', a quantum bit can be represented as a point on a sphere (the Bloch sphere), where the poles correspond to the classical “0” and “1” states, and all other points on the sphere represent coherent superpositions, or combinations, of these states. Single-qubit operations correspond to rotations on the Bloch sphere, and two-qubit operations can result in entanglement between different qubits. Quantum algorithms exploit quantum parallelism and quantum interference to efficiently solve certain problems that are classically intractable. 

A qubit can be made from any two-level system, provided certain criteria are met. In 1996, David DiVincenzo published a paper listing the minimum requirements a candidate technology must have for building a quantum computer \cite{DiVincenzo_1996}. These have come to be known as the ``DiVincenzo criteria.'' They include scalability in the number of qubits, the ability to perform operations and measurements on qubits, and the requirement that the decoherence time, a measure of the lifetime of the quantum information, be much longer than the time to perform operations. The latter two requirements can be in conflict; ensuring that the lifetime of quantum information in the system is long requires {\it isolating} the qubit from its environment, while the need to operate on qubits requires some degree of coupling to the outside world. 

There are several modalities currently being explored for quantum computing, including photonic, atomic, and solid-state systems. This article focuses on solid-state qubits, such as those formed by an electron spin \cite{Loss_1998} or a superconducting circuit \cite{Nakamura_1999}, which have transition frequencies in the GHz range. Because their frequencies are in this range, they are compatible with commercial-off-the-shelf (COTS) RF and microwave components. Solid-state qubits have the further advantage that they leverage decades of investment in fabrication by the semiconductor industry, which also provides a path towards fulfulling the scalability requirement. 

Although convenient from the point of view of using COTS equipment, operation in the few-GHz range introduces the need for cooling the devices to avoid the qubit being inadvertently excited from ``0'' to ``1'' due to thermal effects. The equivalent temperature corresponding to 5 GHz is approximately $T=\frac{hf}{k_b} = 250~mK$, where $h$ is Planck's constant and $k_B$ is Boltzmann's constant. It is necessary to be well below this temperature to avoid thermal excitation, typically \textasciitilde $10~mK$. This is well within the capabilities of dilution refrigerators, which operate by evaporative cooling of a mixture of two isotopes of helium. Thirty years ago, dilution refrigerators were highly specialized products, and operating them required cryogenic training and consumable liquid cryogens for the first stages of cooling. Today, commercial systems generally use pulse-tube coolers that run off wall-plug power. Automated gas handling systems have further transformed dilution refrigerators from specialized equipment to nearly push-button systems. However, understanding the limitations of operating in a cryogenic environment is still important. For example, dilution refrigerators are limited in the cooling power they can provide, and every cable that enters the fridge introduces a passive heat load, simply by being connected to room temperature. If resistive current-carrying wires are used, this can further introduce power that must be dissipated. Additionally, the differential thermal contraction of different materials must be considered when designing low-temperature apparatus, and samples must be properly heat sunk so that they can thermalize with the low-temperature stage of the refrigerator.

Packaging of solid-state qubits in a cryogenic environment introduces challenges due to the sensitivity of the qubits to the electromagnetic environment. For example, it has long been known that at low temperatures, disordered solids have defects that can be modeled as a collection of a large number of two-level systems (TLS) \cite{AHV_1972, Phillips1972}. Electrically-active TLS, such as those found in dielectric materials or at surfaces,  can interact with a qubit's electric field and provide a decay channel for the quantum information \cite{mrsbulletin_2013}. Therefore, it is important to make sure that the fraction of the qubit's electric field that interacts with TLS is small. Similarly, the fraction of the electric field that interacts with normal metal must be small to avoid resistive losses. Magnetic impurities can also impact qubit performance by introducing magnetic field noise, or ``flux noise,'' that can cause decoherence in superconducting qubits. Spurious modes in the packaging, or even on or within the qubit chips themselves, can impact qubit performance by providing a way for quantum information to leave the system in an uncontrolled manner. Worse, coupling to spurious modes can increase microwave crosstalk between qubits and introduce correlated errors, which are particularly hard to correct using quantum error correction schemes. Microwave crosstalk can also result in the control signal for one qubit shifting the energy levels of another qubit through the Stark shift, another source of decoherence.

Developing a packaging scheme that meets all of the requirements for operation of solid-state qubits in a cryogenic environment can be a formidable challenge. In this article, we discuss work being done in our group as well as in the broader community, focusing on the role of 3D integration and packaging in quantum processing with solid-state qubits.  In sections \ref{sec:Tyranny} and \ref{sec:Tailoring}, we discuss the role of 3D integration in building arrays of qubits and controlling their microwave environment. While this discussion focuses on applications to superconducting qubits, we note here that the work is more generally relevant to any solid-state qubit operating at cryogenic temperatures since it is often desirable to use superconducting circuits for readout and control of any qubit in a dilution refrigerator. Sections \ref{sec:GettingOffChip} and \ref{sec:GettingOutOfFridge} focus on the challenges of routing signals from the chip to the output of a dilution refrigerator, and section \ref{sec:Future} discusses challenges and ideas for future systems with larger numbers of qubits.

\section{The Tyranny of Interconnects}\label{sec:Tyranny}

One of the advantages of solid state qubits is that they are amenable to fabrication on silicon wafers with industry-standard processes and tools. While this greatly simplifies fabrication, laterally addressing large numbers of qubits from the edge of a chip quickly becomes infeasible due to interconnect crowding. Just as the semiconductor and imaging industries moved to 3D integration to reduce latency, dissipate power more effeciently, and allow for heterogeneous integration, the solid-state qubit field is also looking to 3D integration to address arrays of coherent qubits, which are needed for some quantum error correction schemes \cite{Fowler_2012}. Solving the interconnect problem to address a 2D array requires ``breaking the plane'' to route wires past each other so they can access the interior qubits. This can be accomplished using either multi-layer fabrication processes or plane-breaking packaging schemes. In this section, we will discuss the former; the latter is covered in section \ref{sec:GettingOffChip}.

\begin{figure}[b]
\includegraphics[width = 5 in]{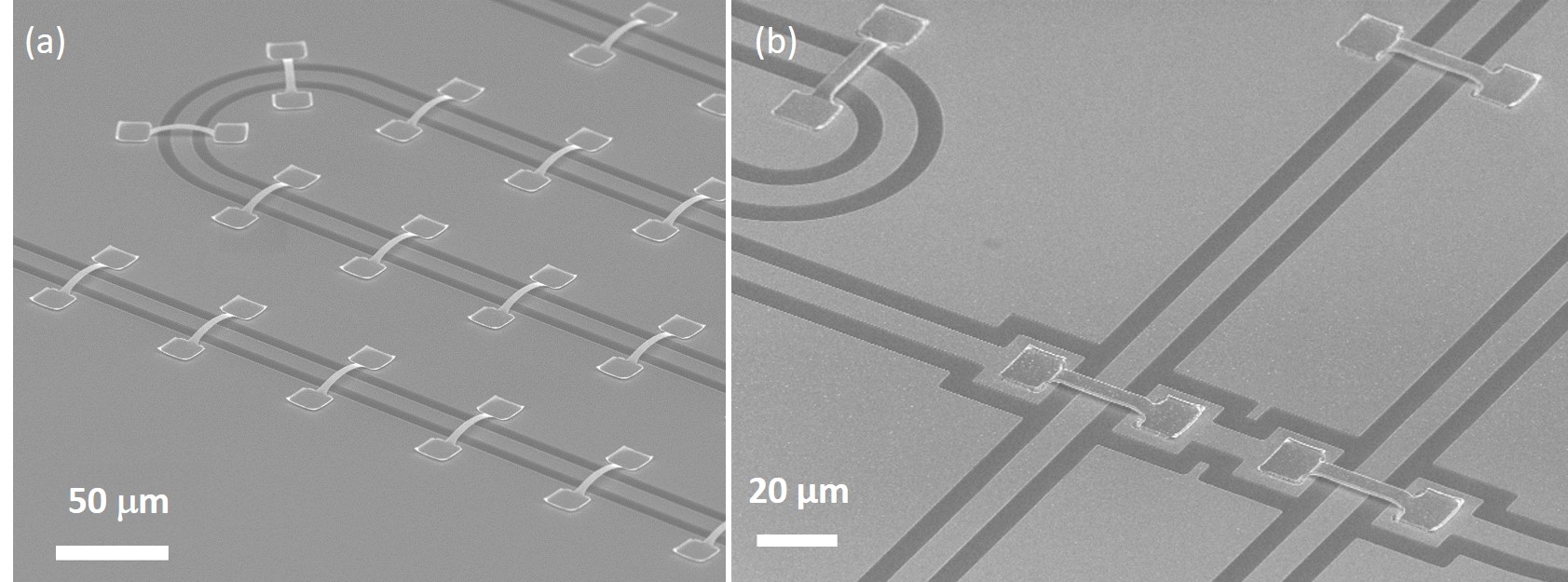}
\caption{\label{fig:airbridges}  Images of air bridges fabricated at MIT Lincoln Laboratory. In (a), the air bridges are used to connect the ground planes on two sides of a coplanar waveguide transmission line resonator, and in (b) the air bridges are used to route coplanar waveguide lines past each other \cite{Yoder}.}
\end{figure}

One multi-layer fabrication method for routing traces past each other is the use of superconducting air bridges, using resist or dielectric material as scaffolding that is removed to leave a free-standing bridge \cite{Chen_2014}. These air bridges can also be used to tie sections of the ground plane together, improving microwave hygiene and reducing the probability of exciting spurious modes. Figure \ref{fig:airbridges} shows images of air bridges fabricated at Lincoln Laboratory to connect ground planes and to route wires past each other. \cite{Yoder}.

\begin{figure*}[htb]
\includegraphics[width = 6 in]{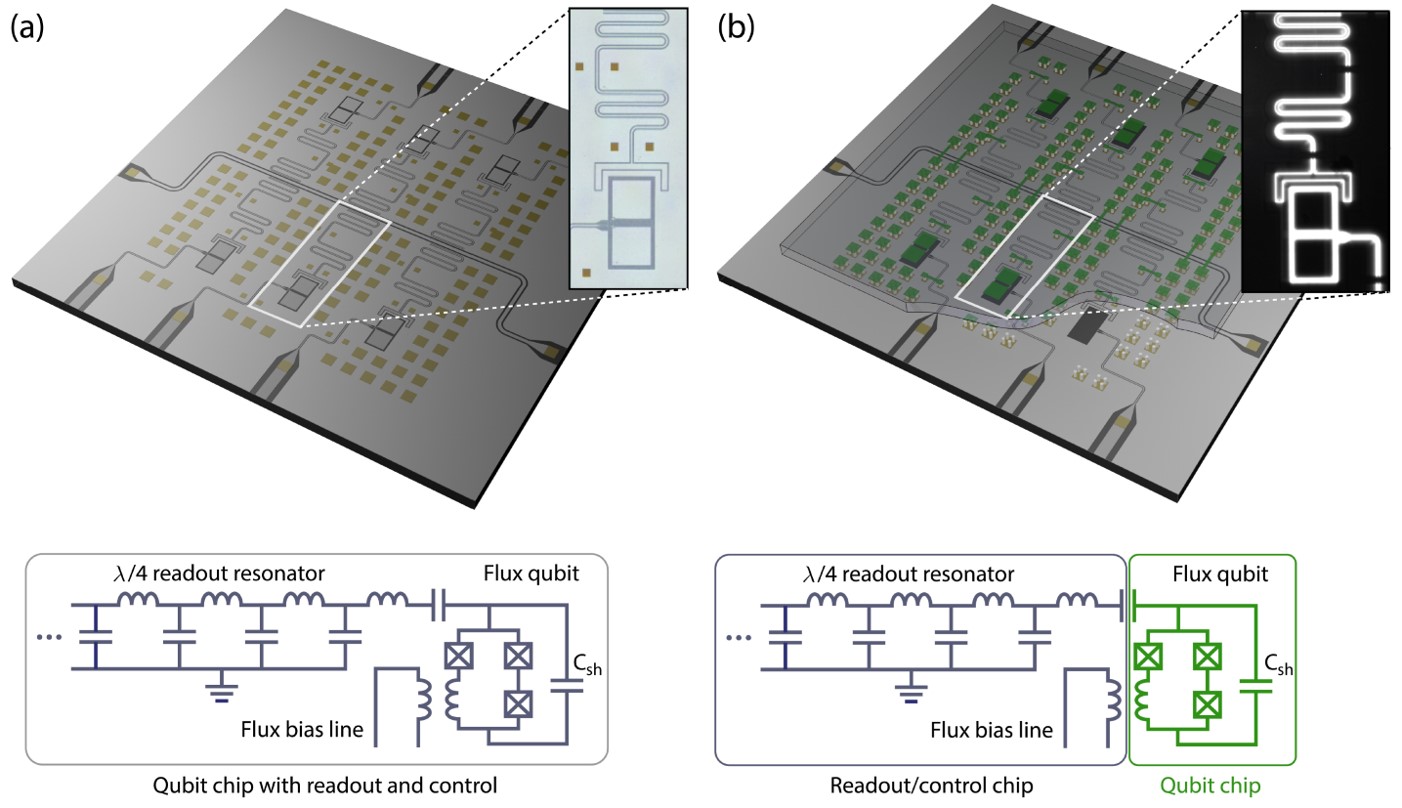}
\caption{\label{fig:flipchip} Comparison of single-chip and flip-chip versions of a superconducting flux qubit. a) Single chip with 6 flux qubits, where each qubit is controlled and read out with on-chip elements. b) 3D integrated qubits, where a chip with six qubits is flipped onto and bonded to another chip with control and readout elements. Qubits on both chips had similar performance. Reprinted from \cite{Rosenberg_2017}.}
\end{figure*}

Another option that enables wires to cross is the use of flip chip integration to bond the qubit chip to another chip. We have used this method to demonstrate off-chip readout and control of a superconducting flux qubit, as shown in Figure \ref{fig:flipchip}.  The left side of the figure shows the layout of a test chip with six superconducting qubits. Each of the qubits has a bias line that applies magnetic flux to the qubit loop, changing its energy, and each has a far-detuned quarter wave transmission line resonator that experiences a shift in resonance based on the qubit state and is used to read out the qubit state. The six readout resonators, which each have a slightly different resonance frequency, are coupled to a single transmission line to enable multiplexed readout. The right side of the figure shows the flip chip version of the same circuit, where the qubits and the control/readout circuitry are split into two different chips. The qubits are on the top chip, which is attached to the bottom control/readout chip using thermocompression bonding of indium bumps. The bottom chip contains the bias lines, which couple inductively across the few-$\mu m$ gap between the chips, and readout resonators, which couple capacitively across the gap. An underbump metallization layer of Ti/Pt/Au provides a low-resistance galvanic path between the chips, which is used to connect the ground plane of the chips. We found that the single-chip qubits had nearly identical lifetimes compared to the flip-chip version, indicating that flip chip processing and bump bonding did not have a significant impact on qubit lifetimes of $\approx 20~\mu s$ \cite{Rosenberg_2017}. Other groups are pursuing flip chip integration, with recent demonstrations of a fully superconducting path between two bump-bonded chips \cite{Foxen_2018,Obrien_2017}.

\begin{figure*}[htb]
\includegraphics[width = 6 in]{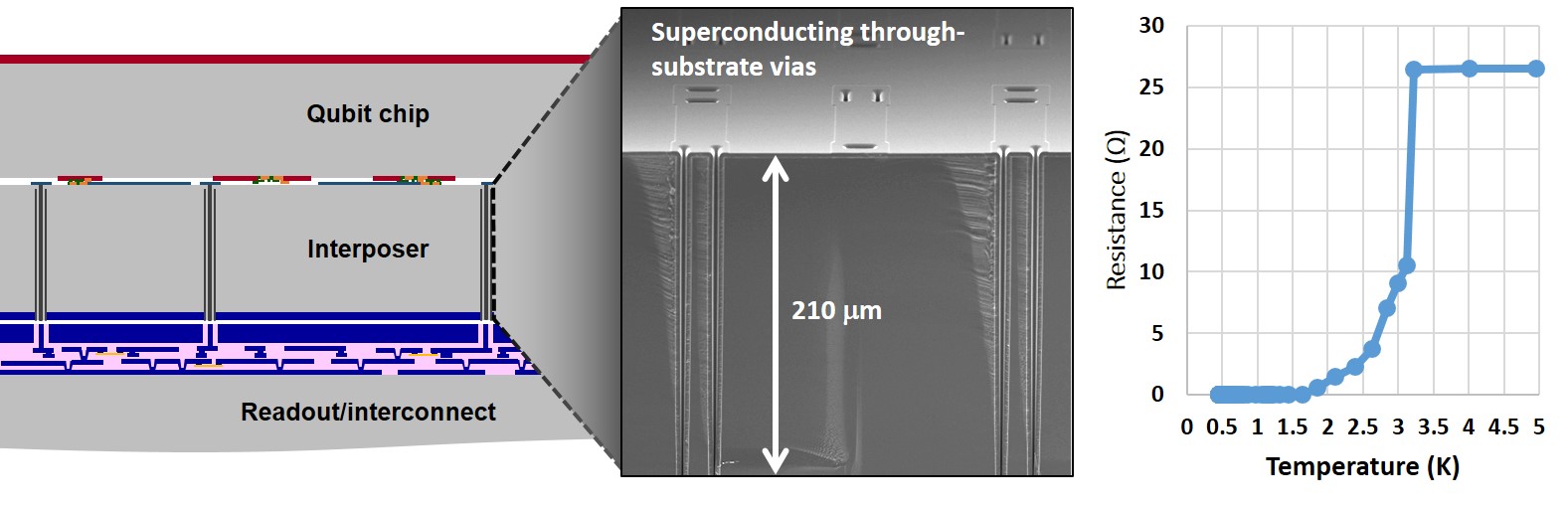}
\caption{\label{fig:ThreeStack}  Envisioned three-tier stack for control and readout of a 3D integrated quantum processor. The three chips are bonded together using indium bump bonds, and superconducting through-substrate vias provide connectivity between the qubit chip on the top and the readout/interconnect chip on the bottom. The inset shows a cross section of a chip with superconducting through-substrate vias. The graph on the right displays a measurement of via resistance as it is cooled through the superconducting  transition temperature.}
\end{figure*}

Although both flip chip integration and superconducting air bridges can be used to access the interior qubits of a 2D array, efficient wire routing requires more layers of metal. The typical method for doing this is to use a multi-layer process, where deposited dielectrics isolate wiring layers. Superconducting multilayer processing has been developed for classical digital superconducting logic at commercial and government foundries \cite{hypres, Tolpygo_2015}. Unfortunately, the lossy dielectric materials generally associated with these processes are not compatible with low-loss superconducting circuits. As a result, simply constructing superconducting microwave circuits on top of a multilayer chip, or bump bonding a qubit chip to a multilayer chip in close proximity, can reduce the lifetime of a solid-state qubit. One approach to solve this problem is to develop multilayer superconducting wiring without lossy materials \cite{Dunsworth_2018}. Another approach, which we are pursuing, is to separate the lossy multilayer chip from the qubit chip through the use of an interposer chip with superconducting through-substrate vias (TSVs), as shown in Figure \ref{fig:ThreeStack} \cite{Rosenberg_2017}. Several groups are developing superconducting TSVs for vertical signal delivery and for reducing spurious chip modes \cite{Vahidpour_2017,Versluis_2016}. 

Our approach to TSV fabrication focuses on the development of compact, high-aspect-ratio TSVs to enable high-density vertical wiring. We first use reactive ion etching to etch $200~\mu m$-deep blind vias in a $725~\mu m$ thick, $200~mm$ diameter silicon wafer. After the vias have been etched, we deposit superconducting TiN on the wafer and line the TSVs using chemical vapor deposition. The planar metal is patterned and the wafer is then flipped and temporarily bonded to a carrier wafer. The TSVs are revealed by using chemical mechanical planarization to thin the wafer to $200~\mu m$. Superconducting metal is deposited and patterned on the revealed side of the TSV wafer, and the wafer is diced and the chips are debonded from the carrier. The resulting chip has superconducting metal on both sides, with connections between the top and bottom plane provided by compact ($10~\mu m$ x $25~\mu m$) superconducting TSVs. Figure \ref{fig:ThreeStack} shows a scanning electron micrograph of a cross-sectioned TSV chip, showing the superconductor-lined TSVs and patterned metal on one side of the chip. 

We have characterized the superconducting TSVs at DC and microwave frequencies. At DC frequencies, we perform 4-wire resistance measurements on test structures comprising many TSV links in series, connected to each other through alternating strips of metal on the top and bottom of the chip. These measurements indicate excellent yield; chains with as many as 3,200 TSV links in series have displayed a superconducting critical temperature of 2.5-3 Kelvin \cite{Mallek}.

\begin{figure}[hb]
\includegraphics[width = 4 in]{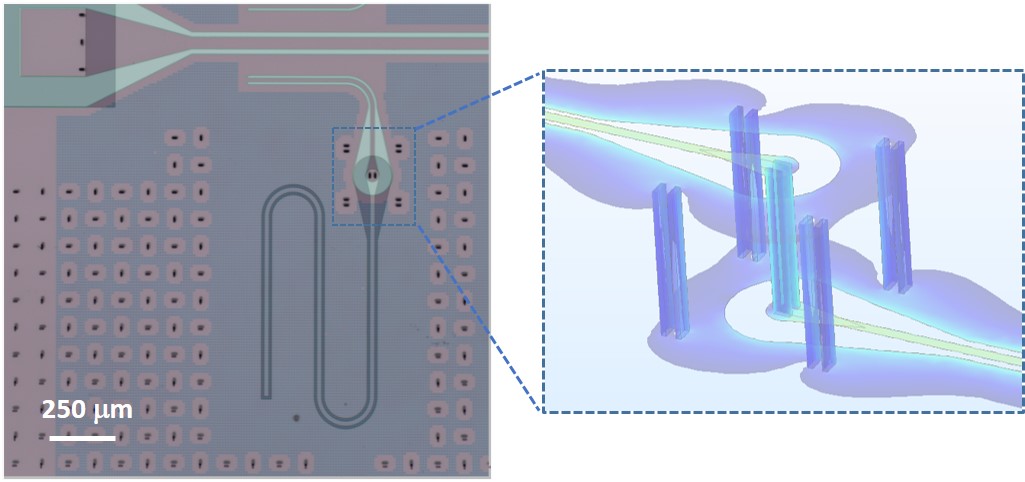}
\caption{\label{fig:uwaveTSV} Example of a shielded microwave TSV transition designed to minimize reflections. The main figure shows the top and bottom of a TSV chip overlaid, with the bottom image mirrored for ease of viewing. The structure shown is a quarter-wave coplanar waveguide transmission line resonator that is split between the top and bottom of the chip, with a microwave TSV transition connecting them. A pair of TSVs connect the center traces on either side of the chip, and four pairs of TSVs connect the ground planes. The inset shows the results of a simulation of the current flow in the transition.}
\end{figure}

At microwave frequencies, a single TSV connection in the signal line of a planar waveguide structure would introduce a significant impedance mismatch between high-frequency lines on either side of the chip, resulting in signal reflection and distortion. Therefore, it is essential to design a TSV transition that minimizes such mismatch and properly routes the return currents for microwave lines. The inset of Figure \ref{fig:uwaveTSV} shows one of the structures we have designed to connect co-planar waveguide (cpw) transmission lines on two sides of a chip. A pair of TSVs provides a connection for the center trace of the cpw, and four pairs of TSVs connect the ground plane on either side of the chip. The cpw shape near the TSV transition is tailored to minimize reflections of microwave signals, and the simulated reflection at the transition is less than -30 dB. 

In order to characterize the losses associated with the TSV transition, we incorporated the structure shown in the inset to Figure \ref{fig:uwaveTSV} into superconducting transmission line resonators. As shown in Figure \ref{fig:uwaveTSV}, each resonator comprises two sections of transmission lines, one on the top of the chip and one on the bottom, with a TSV transition providing a connection between the two sections. Measuring the quality factor of the resonator can indicate if the TSV transition is introducing loss at microwave frequencies.  In most microfabricated superconducting resonators, the quality factor at low power is limited by the presence of two-level systems at interfaces that interact with the electric field and consitute a loss mechanism \cite{Martinis_2005}. The effect of these two-level systems is dependent on temperature and power, so it is essential to measure the quality factor at low temperatures and at a sufficiently low power that the mean photon number in the resonator is approximately unity, to reproduce the environment seen by the qubits. If we are interested in characterizing the loss associated with a particular component integrated into a resonator, such as a microwave TSV transition, shifting the location of the component relative to the voltage and current nodes of the resonator can provide useful insight; if the primary loss mechanism is resistive, it will have the biggest effect at the current anti-node, and if it is related to two-level systems, the effect will be largest at the voltage anti-node where the electric field is high. We have measured quality factors of more than 20 resonators of the type shown in Fig \ref{fig:uwaveTSV}, with locations of the transition ranging from the position shown in the figure (near the current anti-node) to the voltage anti-node. In all cases, the quality factor ranged from $100,000$ to $300,000$, consistent with the intrinsic quality factors of planar transmission line resonators on each surface \cite{Yost}. These results indicate that the TSVs do not introduce significant ohmic or TLS loss at microwave frequencies.

\section{Using 3D integration to tailor the microwave environment}\label{sec:Tailoring}

In addition to solving the interconnect problem, 3D integration can also be used to tightly control the microwave environment surrounding the qubit to an unprecedented degree. The microwave environment surrounding the qubits and control/readout lines can influence the performance of a quantum processor because any mode that couples to the qubit can reduce the qubit lifetime and/or result in microwave crosstalk. Of course, some interaction with the outside world is needed in order to control and read out the qubits, and quantum circuit designers must carefully balance the need for control of the quantum system with the potential for adding loss channels. 

Spurious modes that can impact performance include cavity resonances, quasi-lumped-element modes involving the inductance and capacitance between the chip and package grounds, and modes corresponding to undesired excitations of superconducting planar elements. Box modes can be excited in any cavity with metallized walls, at a frequency that depends on the shape and size of the cavity. In general, for packaged single-chip superconducting circuits on silicon chips, two of the dimensions are much larger than the third, so the lowest-frequency mode that is excited is the transverse magnetic TM110 mode. If the dielectric in the cavity is air, a cavity with the largest dimensions 10 mm x 10 mm will have a resonant frequency of 22 GHz, while for a 10x10 mm silicon chip with $\epsilon_r  \approx 11.5$, the lowest box mode frequency is 6 GHz. Within the qubit package, care must be taken to avoid large cavities that can host box modes, or the qubit must be shielded from them. Box modes in the chip can be mitigated by including an air-gap under the chip to increase the frequencies of the modes, but as chips get larger, eventually another approach must be used, such as the use of vias to decouple the frequency dependence of the chip box modes from the size of the chip \cite{Vahidpour_2017}.

The finite-impedance connections between the chip ground plane and the package ground are another source of spurious modes. Researchers have shown that the inductance of the ground plane wire bonds, when combined with the capacitance between the chip ground plane the package ground plane, can result in resonances near qubit frequencies and significantly increase microwave crosstalk \cite{Wenner_2011}. One method for reducing the impact of these modes is to reduce the capacitance between the chip ground plane and the package ground, which can push the resonances higher than the frequency range of interest. This can be accomplished by selectively removing metal directly below the qubit chip \cite{Wenner_2011}, or by using TSVs to directly short out the capacitance. Similarly, flip chip bump bonds can be used instead of wire bonds to significantly reduce the inductance of the connection between the chip and package ground planes.

Finally, spurious modes can result from the excitation of alternate waveguide modes in the superconducting metal than the ones intended. For example, slotline modes can be excited on a coplanar waveguide if there is not a good connection between the groundplane on both sides of the signal trace, or if there are discontinuities in the ground plane. The probability of exciting these modes can be greatly reduced by forming a well-connected ground plane, which can be accomplished using superconducting bump bonds in a flip chip configuration \cite{Foxen_2018}, or by using superconducting air bridges \cite{Dunsworth_2018}.

\begin{figure}[b]
\includegraphics[width = 3 in]{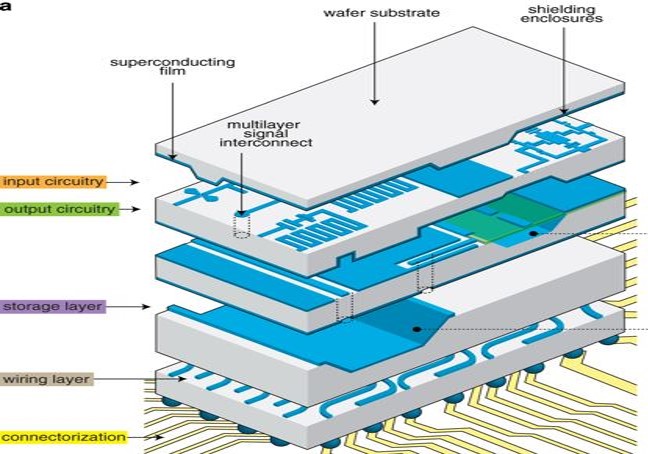}
\caption{\label{fig:MMIQC} Concept for a multilayer microwave circuit, reprinted from \cite{Brecht_2016}. Wafers are micromachined and bonded together with superconducting materials to form shielding enclosures and circuitry.}
\end{figure}

In many cases, finite-element electromagnetic simulation software can be used to identify, determine the impact of, and mitigate the presence of spurious modes. For superconducting circuits, a technique called black-box quantization \cite{BBQ, Brune} provides a useful framework for determining how spurious modes and coupling can impact qubit performance. Classical simulation software is first used to find the admittance that the qubit sees at microwave frequencies. Foster’s method or Brune synthesis \cite{Brune} is then used to provide an equivalent lumped-element circuit that reproduces the results of the simulation, and those lumped elements can then be incorporated into the circuit model to determine the effect on the qubit. 

Exciting recent work has focused on the use of 3D integration to build enclosures around qubits and components to shield them from unintended coupling to other components or modes. For example, researchers have created a micromachined superconducting ``cap'' that is bump bonded to a superconducting qubit chip to isolate the qubit from other elements in the circuit. The cap has a layer of sputtered aluminum lined with a continuous layer of thin molybdenum to prevent native oxidation of the aluminum, and indium bumps provide a connection between elements on the two chips \cite{Obrien_2017}.  In other work, researchers are looking at micromachined superconducting cavities not only for shielding, but also to store and manipulate quantum information. Figure \ref{fig:MMIQC} shows a concept for a multilayer microwave integrated quantum circuit (MMIQC) \cite{Brecht_2016}, which includes vertical interconnects, shielding enclosures, and high-Q storage cavities. Some elements of this concept, such as coupling superconducting qubits to high-Q micromachined cavities, have been demonstrated \cite{Brecht_2017}.

\section{Getting signals off the chip}\label{sec:GettingOffChip}

\begin{figure*}[hbtp]
\includegraphics[width = 5 in]{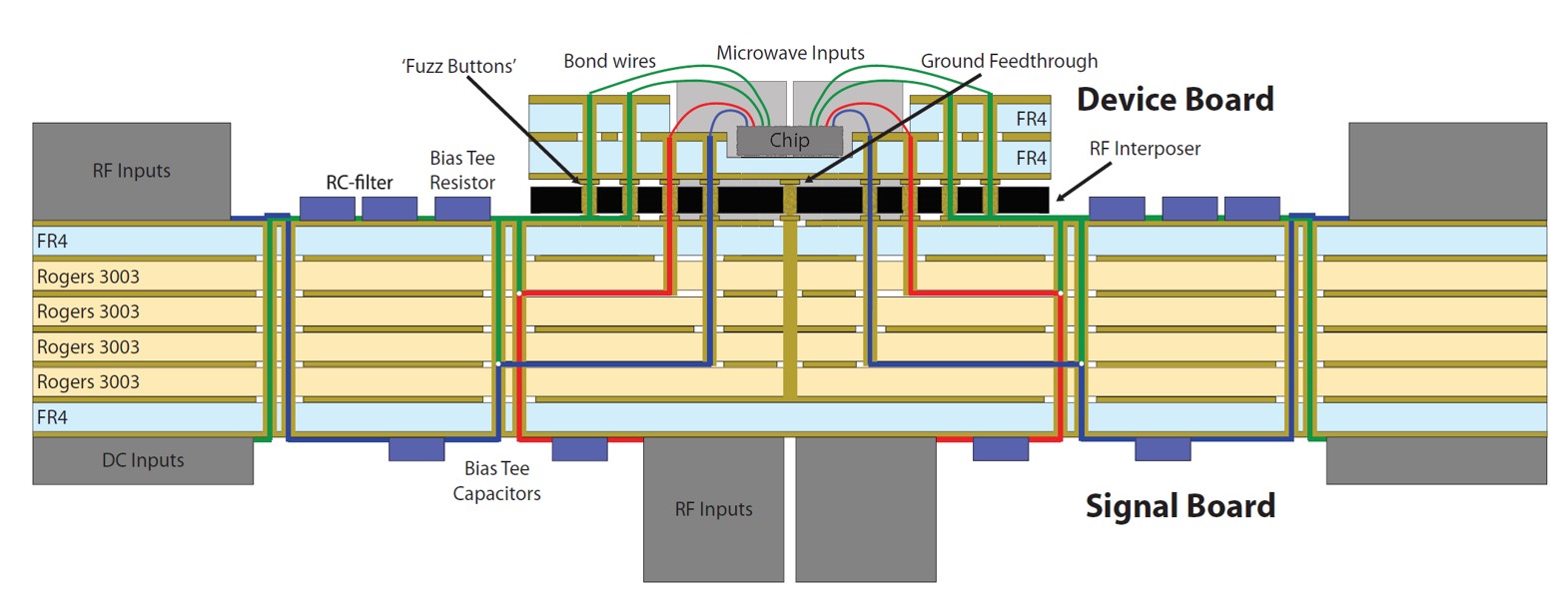}
\caption{\label{fig:CryoInterconnect} Interconnect platform developed for controlling and reading out solid-state qubits. The qubit chip is bonded to a small device board, which is then connected to a more complex, reusable board through contact pins. Reprinted from \cite{Colless_2014}}
\end{figure*}

Whether a single chip or a 3D-integrated chip is used, there must be a reliable interface between the silicon and a path that eventually leads out of the fridge to room temperature. This connection can be made either directly to the qubit chip or through a printed circuit board (PCB) or other interposer that can be interfaced with microwave connectors. The standard method is to wire bond between the edges of a silicon chip and a PCB. In some cases, the PCB may house components that are needed for qubit operation. For example, Figure \ref{fig:CryoInterconnect} shows a schematic of a cryogenic package developed for solid-state qubits. The authors' modular platform, which includes 74 DC connections and 36 RF and microwave connections, comprises two PCBs: a simpler one that is wire-bonded directly to the qubit chip, and a more complex board that includes components such as bias tees and filters. With this system, the cost and complexity are mainly in the larger board, so a qubit chip can be permanently bonded to the smaller board \cite{Colless_2014}.

\begin{figure}[hbtp]
\includegraphics[width = 5 in]{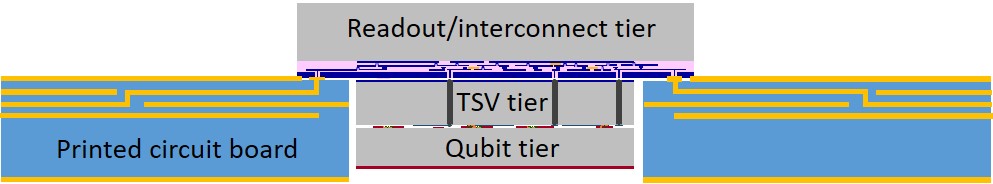}
\caption{\label{fig:3stackInt} Schematic of the three-tier stack shown in Figure \ref{fig:ThreeStack} attached via flip chip to a microwave interposer. The readout and interconnect chip is larger than the other chips, allowing a peripheral connection to the microwave interposer. Depending on the density of wires, one or more rows of bumps can make a connection to buried striplines in the PCB.}
\end{figure}

Flip chip bump bonds can also be used to connect solid state qubits to a microwave interposer \cite{Caudillo_2018}. For example, we are investigating using indium or solder bumps to form the connection between the three-tier stack discussed in Section \ref{sec:Tyranny} and a PCB, as shown in Figure \ref{fig:3stackInt}. As discussed in Section \ref{sec:Tailoring}, the use of bumps rather than wire bonds is expected to provide better impedance matching and reduce crosstalk. An important consideration is the coefficient of thermal expansion for the silicon chip and the microwave interposer, which should be adequately matched to ensure mechanical robustness with thermal cycling. 

Some researchers have noted that interfacing with a PCB or connectors offers another opportunity for ``breaking the plane'' to allow access to interior qubits in a 2D array. For example, the authors of Ref \cite{Bronn_2018} use pogo pins to make contact between a PCB and an array of seven qubits on a silicon chip with an interior qubit that cannot be addressed laterally, and the authors of Ref \cite{Rahanim_2017, Patterson_2019} use a novel double-sided coaxial line approach to capacitively couple to qubits for control and read out. Researchers have also flipped a silicon chip with 13 qubits directly on top of a PCB, making contact by pushing the silicon chip against vias in the PCB signal lines \cite{Liu_2017}. Finally, a concept called the ``quantum socket'' uses custom spring-loaded micro-wires to make contact between pads on the silicon chip and standard microwave connectors \cite{Bejanin_2016}.

These novel approaches greatly simplify testing and prototyping by shifting the wiring complexity to the PCB, where well-established techniques exist to route transmission lines within multiple layers. However, there are some drawbacks to approaches that require directly contacting to a qubit chip. First, though photolithographic techniques used on silicon chips generally can define very small features $\approx~0.1-1~\mu m$, printed circuit boards and pins have much larger features, typically $> 100~\mu m$. This larger feature size and the desire to avoid complicated alignment procedures pushes the required contact pad size on the qubit chip to a few hundred $\mu m$. Contact pads of this size limit the density of active components, and can lead to excess crosstalk. Furthermore, pins and PCBs generally contain dielectric layers that could lead to reductions in qubit coherence time, so care must be taken to spatially isolate the qubits from these materials. Finally, contact resistance is generally higher for contacts based on pressure rather than wire bonds or bump bonds. This is particularly a problem in schemes that involve directly contacting the qubit chip, since any resistive heating will be generated directly on the qubit chip.

Resistive heating within the interposer can also be an issue when the number of lines in a package becomes high enough. The use of superconducting materials on microwave interposers is of interest both to reduce heating and to improve signal quality. Research has shown that the use of superconducting materials can vastly reduce transients observed when applying fast control pulses to qubits \cite{Foxen_2018_arxiv}. 

\begin{figure*}[htb]
\includegraphics[width = \textwidth]{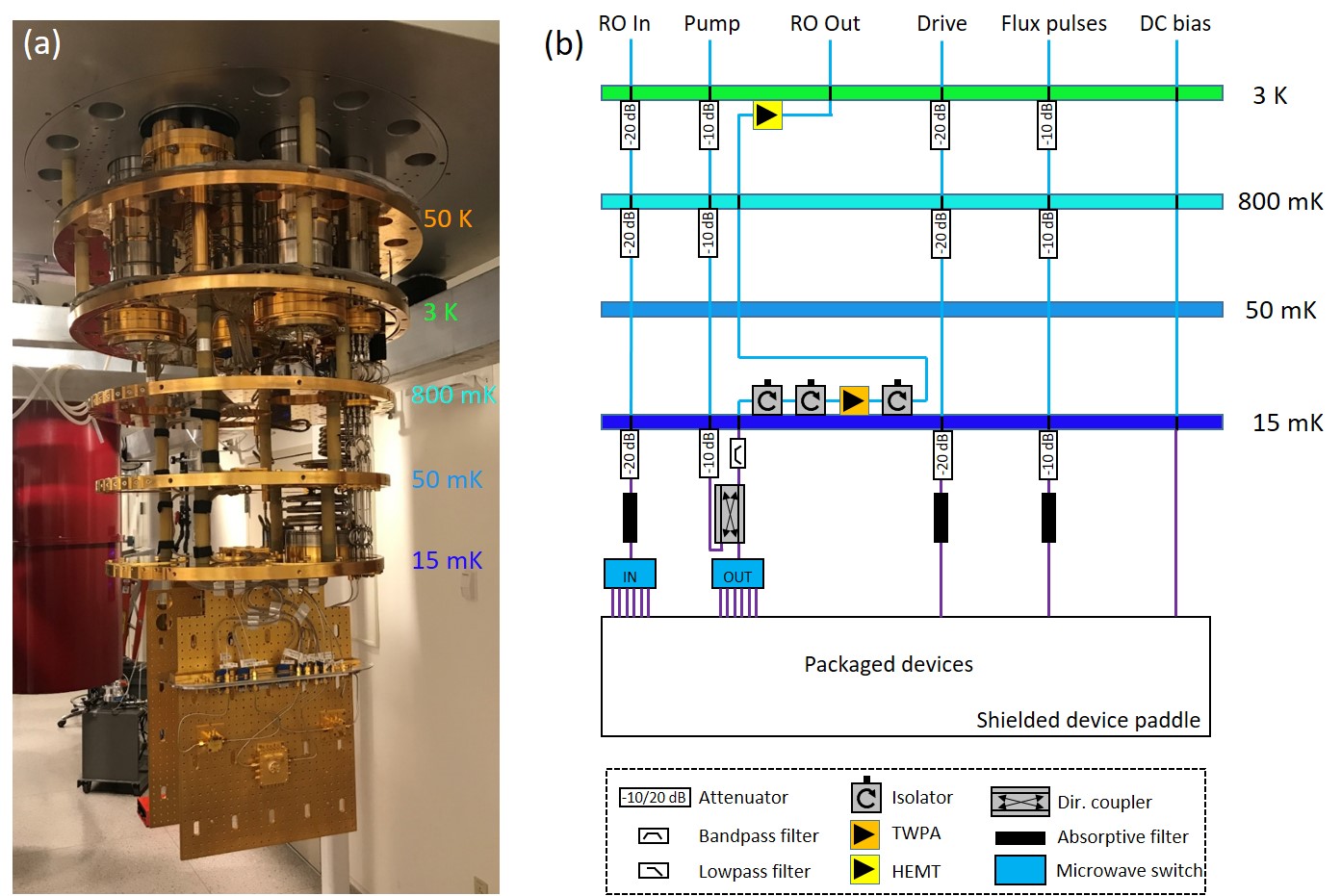}
\caption{\label{fig:FridgeWiring} Dilution fridge wiring for superconducting qubit experiments. (a) Photograph of a dilution refrigerator. Each circular plate is operated at a progressively lower temperature. (b) Example schematic of the internal fridge wiring for typical superconducting qubit experiments.  Microwave cables are used for qubit readout, amplifier pump tones, qubit drive tones, and magnetic flux pulses.}
\end{figure*}

\section{Getting out of the fridge}\label{sec:GettingOutOfFridge}

Now that we have discussed strategies for routing signals on and off of the qubit chip, we also need to consider how to get signals into and out of the dilution fridge.  Fridge wiring is typically divided into two categories-- RF and DC.  The RF wiring is typically used to control and read out the qubits, while DC wiring can be used for qubit control and for powering active components inside the fridge, such as amplifiers and switches.  Due to the cryogenic environment and the high sensitivity of the qubits to noise, several mechanical and electrical considerations must be taken into account when wiring a dilution fridge for qubit experiments.  

As shown in Figure \ref{fig:FridgeWiring}, the inside of a dilution fridge is composed of stages that are operated at progressively lower temperatures.  In order for each stage to remain cold, it needs to be thermally isolated from the warmer stages above.  The fridge wiring acts as a heat link between stages, creating a passive heat load that scales with the number of connections.  Therefore, when scaling up to larger numbers of connections, careful attention must be paid to thermal engineering, as described in Ref \cite{Krinner_2018}.  For example, typically materials with low thermal conductivity are used, such as stainless steel and superconducting metals.  It is also important to minimize the active heat load from resistive losses in the DC lines by reducing contact resistances and using superconducting wires whenever possible.  

While these mechanical wiring considerations affect the fridge temperature, electrical considerations can have a more direct impact on qubit performance.  Because solid-state qubits are highly sensitive to noise, careful steps must be taken to protect the qubits from room temperature thermal noise and any added noise from the control electronics.  This is achieved, in part, by applying low-pass filters to the DC lines and progressively attenuating the RF lines at several different temperature stages.  In addition, IR absorbing shields and filters are often used at the coldest temperature stage to protect the qubits from radiation coming from the warmer stages \cite{Barends_2011, Corcoles_2011}.

Finally, there are additional considerations that come into play when designing the microwave readout chain.  Solid-state qubits are often read out through capacitively coupled microwave resonators.  The resonators are designed to efficiently measure the qubit state when probed at very low powers, on the order of -110 dBm or lower. To achieve high fidelity readout, these small signals must then be amplified with as little added noise as possible.  Most state-of-the-art readout chains use nearly-quantum-limited amplifiers, such as the traveling-wave parametric amplifier (TWPA) \cite{Macklin_2015} at the lowest temperature stage and high-electron-mobility transistor (HEMT) amplifiers at the 3 Kelvin stage.  

Looking forward, considerable engineering efforts are needed to scale up existing infrastructure to accommodate larger solid-state qubit processors.  Existing approaches to fridge wiring will be difficult to extend beyond the scale of hundreds of qubits due to thermal and space constraints.  Several alternative approaches are currently under consideration, including commercial multi-coax assemblies and superconducting flex-print \cite{Tuckerman_2016,Walter_2018} and rigid-flex wiring \cite{Das_2018}.  There are also efforts underway to develop on-chip microwave isolators to replace existing bulky components made with magnetic materials \cite{Kamal_2014, Chapman_2018}.

\section{Looking to the future}\label{sec:Future}

\begin{figure}[h]
\includegraphics[width = 3 in]{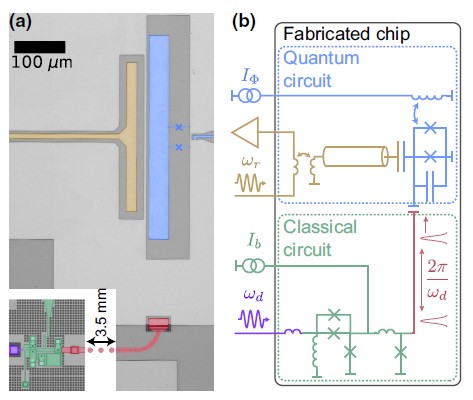}
\caption{\label{fig:SFQ} Single flux quantum (SFQ) circuit coupled to qubit, reprinted from \cite{Leonard_2019}. The SFQ driver circuit delivers a train of pulses at a subharmonic of the qubit frequency to perform single-qubit control. }
\end{figure}

As the number of qubits in a system gets larger, the brute force approach of generating control signals at room temperature and sending one or more lines into the cryostat for each qubit will no longer be feasible. It has been estimated \cite{NAP25196} that this will occur at approximately 1,000 physical qubits. Several options exist for generating control signals at cryogenic temperatures, including cryogenic CMOS and superconducting digital electronics such as single-flux-quantum-based (SFQ) circuits. Researchers have demonstrated the use of cryogenic CMOS \cite{Bardin_2019} and SFQ \cite{Kirill_2014,Leonard_2019} circuits for readout and control of superconducting qubits. Figure \ref{fig:SFQ} shows a circuit schematic from Ref \cite{Leonard_2019} for an SFQ driver used for coherent single-qubit control. Single-qubit gate fidelities as high as 95\% were observed, with the primary limitation being from non-equilibrium quasiparticles generated in the SFQ driver. The authors suggest several options for mitigating this, including the use of flip-chip bonding to separate the SFQ chip from the qubit chip. It is estimated that addressing the quasiparticle poisoning problem can increase the gate fidelity past the fault-tolerant threshold \cite{McDermott_2014, Liebermann_2016}.

A complementary approach to generating signals at cryogenic temperatures is to convert to the optical domain and utilize the large bandwidth available on optical fiber. The process of converting between the microwave and optical domain can be performed classically, for example by mixing two optical signals on a photo-detector, or in a way that preserves the quantum information in a single photon. If the quantum information is preserved, a bi-directional microwave to optical interface would also enable an optical link to connect nodes of a quantum computer in a distributed quantum computing environment, and it could form an integral part of a quantum network \cite{Kimble_2018}. Coherent microwave to optical conversion has been demonstrated using Rydberg atoms \cite{Han_2018}  with 0.3\% efficiency, and with 9\% in a system using a mechanical resonator to mediate the microwave to optical interaction \cite{Andrews_2014}. In both cases, there are clear paths to increase the efficiency, though there are still many integration challenges to overcome with this approach. 

The nascent field of quantum engineering is steadily advancing towards commercial viability. While there are still numerous and significant technical hurdles to overcome before a fault-tolerant quantum computer can be built, we are currently at the stage where an interdisciplinary approach is needed. In particular, microwave engineering and packaging are central to the ability to develop more complex systems, and the field will benefit from more collaborations between scientist and engineers in this area. 

\acknowledgements{
We gratefully acknowledge E. Golden, X. Miloshi, P. Murphy and A. Sevi for technical assistance, and J. Cummings, E. Dauler, J. Rokosz, and R. Slattery for useful discussions. This research was funded by the Office of the Director of National Intelligence (ODNI), Intelligence Advanced Research Projects Activity (IARPA) and by the Assistant Secretary of Defense for Research \& Engineering under Air Force Contract No. FA8721-05-C-0002. The views and conclusions contained herein are those of the authors and should not be interpreted as necessarily representing the official policies or endorsements, either expressed or implied, of ODNI, IARPA, or the US Government.}

\bibliography{scqubits}

\begin{thebibliography}{55}%
\makeatletter
\providecommand \@ifxundefined [1]{%
 \@ifx{#1\undefined}
}%
\providecommand \@ifnum [1]{%
 \ifnum #1\expandafter \@firstoftwo
 \else \expandafter \@secondoftwo
 \fi
}%
\providecommand \@ifx [1]{%
 \ifx #1\expandafter \@firstoftwo
 \else \expandafter \@secondoftwo
 \fi
}%
\providecommand \natexlab [1]{#1}%
\providecommand \enquote  [1]{``#1''}%
\providecommand \bibnamefont  [1]{#1}%
\providecommand \bibfnamefont [1]{#1}%
\providecommand \citenamefont [1]{#1}%
\providecommand \href@noop [0]{\@secondoftwo}%
\providecommand \href [0]{\begingroup \@sanitize@url \@href}%
\providecommand \@href[1]{\@@startlink{#1}\@@href}%
\providecommand \@@href[1]{\endgroup#1\@@endlink}%
\providecommand \@sanitize@url [0]{\catcode `\\12\catcode `\$12\catcode
  `\&12\catcode `\#12\catcode `\^12\catcode `\_12\catcode `\%12\relax}%
\providecommand \@@startlink[1]{}%
\providecommand \@@endlink[0]{}%
\providecommand \url  [0]{\begingroup\@sanitize@url \@url }%
\providecommand \@url [1]{\endgroup\@href {#1}{\urlprefix }}%
\providecommand \urlprefix  [0]{URL }%
\providecommand \Eprint [0]{\href }%
\providecommand \doibase [0]{http://dx.doi.org/}%
\providecommand \selectlanguage [0]{\@gobble}%
\providecommand \bibinfo  [0]{\@secondoftwo}%
\providecommand \bibfield  [0]{\@secondoftwo}%
\providecommand \translation [1]{[#1]}%
\providecommand \BibitemOpen [0]{}%
\providecommand \bibitemStop [0]{}%
\providecommand \bibitemNoStop [0]{.\EOS\space}%
\providecommand \EOS [0]{\spacefactor3000\relax}%
\providecommand \BibitemShut  [1]{\csname bibitem#1\endcsname}%
\let\auto@bib@innerbib\@empty
\bibitem [{\citenamefont {Shor}(1997)}]{Shor_1995}%
  \BibitemOpen
  \bibfield  {author} {\bibinfo {author} {\bibfnamefont {P.}~\bibnamefont
  {Shor}},\ }\href {\doibase 10.1137/S0097539795293172} {\bibfield  {journal}
  {\bibinfo  {journal} {SIAM Journal on Computing}\ }\textbf {\bibinfo {volume}
  {26}},\ \bibinfo {pages} {1484} (\bibinfo {year} {1997})},\ \Eprint
  {http://arxiv.org/abs/https://doi.org/10.1137/S0097539795293172}
  {https://doi.org/10.1137/S0097539795293172} \BibitemShut {NoStop}%
\bibitem [{\citenamefont {Biamonte}\ \emph {et~al.}(2017)\citenamefont
  {Biamonte}, \citenamefont {Wittek}, \citenamefont {Pancotti}, \citenamefont
  {Rebentrost}, \citenamefont {Wiebe},\ and\ \citenamefont
  {Lloyd}}]{Biamonte_2017}%
  \BibitemOpen
  \bibfield  {author} {\bibinfo {author} {\bibfnamefont {J.}~\bibnamefont
  {Biamonte}}, \bibinfo {author} {\bibfnamefont {P.}~\bibnamefont {Wittek}},
  \bibinfo {author} {\bibfnamefont {N.}~\bibnamefont {Pancotti}}, \bibinfo
  {author} {\bibfnamefont {P.}~\bibnamefont {Rebentrost}}, \bibinfo {author}
  {\bibfnamefont {N.}~\bibnamefont {Wiebe}}, \ and\ \bibinfo {author}
  {\bibfnamefont {S.}~\bibnamefont {Lloyd}},\ }\href {\doibase
  10.1038/nature23474} {\bibfield  {journal} {\bibinfo  {journal} {Nature}\
  }\textbf {\bibinfo {volume} {549}},\ \bibinfo {pages} {195} (\bibinfo {year}
  {2017})}\BibitemShut {NoStop}%
\bibitem [{\citenamefont {Farhi}\ \emph {et~al.}(2014)\citenamefont {Farhi},
  \citenamefont {Goldstone},\ and\ \citenamefont {Gutman}}]{QAOA}%
  \BibitemOpen
  \bibfield  {author} {\bibinfo {author} {\bibfnamefont {E.}~\bibnamefont
  {Farhi}}, \bibinfo {author} {\bibfnamefont {J.}~\bibnamefont {Goldstone}}, \
  and\ \bibinfo {author} {\bibfnamefont {S.}~\bibnamefont {Gutman}},\
  }\href@noop {} {\enquote {\bibinfo {title} {A quantum approximate
  optimization algorithm},}\ } (\bibinfo {year} {2014}),\ \Eprint
  {http://arxiv.org/abs/1411.4028} {arXiv:1411.4028 [quant-ph]} \BibitemShut
  {NoStop}%
\bibitem [{\citenamefont {Babbush}\ \emph {et~al.}(2018)\citenamefont
  {Babbush}, \citenamefont {Wiebe}, \citenamefont {McClean}, \citenamefont
  {McClain}, \citenamefont {Neven},\ and\ \citenamefont {Chan}}]{qsim}%
  \BibitemOpen
  \bibfield  {author} {\bibinfo {author} {\bibfnamefont {R.}~\bibnamefont
  {Babbush}}, \bibinfo {author} {\bibfnamefont {N.}~\bibnamefont {Wiebe}},
  \bibinfo {author} {\bibfnamefont {J.}~\bibnamefont {McClean}}, \bibinfo
  {author} {\bibfnamefont {J.}~\bibnamefont {McClain}}, \bibinfo {author}
  {\bibfnamefont {H.}~\bibnamefont {Neven}}, \ and\ \bibinfo {author}
  {\bibfnamefont {G.~K.-L.}\ \bibnamefont {Chan}},\ }\href {\doibase
  10.1103/PhysRevX.8.011044} {\bibfield  {journal} {\bibinfo  {journal} {Phys.
  Rev. X}\ }\textbf {\bibinfo {volume} {8}},\ \bibinfo {pages} {011044}
  (\bibinfo {year} {2018})}\BibitemShut {NoStop}%
\bibitem [{\citenamefont {Divincenzo}(1996)}]{DiVincenzo_1996}%
  \BibitemOpen
  \bibfield  {author} {\bibinfo {author} {\bibfnamefont {D.~P.}\ \bibnamefont
  {Divincenzo}},\ }\href@noop {} {\enquote {\bibinfo {title} {Topics in quantum
  computers},}\ } (\bibinfo {year} {1996}),\ \Eprint
  {http://arxiv.org/abs/9612126} {arXiv:9612126 [cond-mat]} \BibitemShut
  {NoStop}%
\bibitem [{\citenamefont {Loss}\ and\ \citenamefont
  {DiVincenzo}(1998)}]{Loss_1998}%
  \BibitemOpen
  \bibfield  {author} {\bibinfo {author} {\bibfnamefont {D.}~\bibnamefont
  {Loss}}\ and\ \bibinfo {author} {\bibfnamefont {D.~P.}\ \bibnamefont
  {DiVincenzo}},\ }\href {\doibase 10.1103/PhysRevA.57.120} {\bibfield
  {journal} {\bibinfo  {journal} {Phys. Rev. A}\ }\textbf {\bibinfo {volume}
  {57}},\ \bibinfo {pages} {120} (\bibinfo {year} {1998})}\BibitemShut
  {NoStop}%
\bibitem [{\citenamefont {Nakamura}\ \emph {et~al.}(1999)\citenamefont
  {Nakamura}, \citenamefont {Pashkin},\ and\ \citenamefont
  {Tsai}}]{Nakamura_1999}%
  \BibitemOpen
  \bibfield  {author} {\bibinfo {author} {\bibfnamefont {Y.}~\bibnamefont
  {Nakamura}}, \bibinfo {author} {\bibfnamefont {Y.~A.}\ \bibnamefont
  {Pashkin}}, \ and\ \bibinfo {author} {\bibfnamefont {J.~S.}\ \bibnamefont
  {Tsai}},\ }\href {\doibase 10.1038/19718} {\bibfield  {journal} {\bibinfo
  {journal} {Nature}\ }\textbf {\bibinfo {volume} {398}},\ \bibinfo {pages}
  {786} (\bibinfo {year} {1999})}\BibitemShut {NoStop}%
\bibitem [{\citenamefont {Anderson}\ \emph {et~al.}(1972)\citenamefont
  {Anderson}, \citenamefont {Halperin},\ and\ \citenamefont
  {Varma}}]{AHV_1972}%
  \BibitemOpen
  \bibfield  {author} {\bibinfo {author} {\bibfnamefont {P.~W.}\ \bibnamefont
  {Anderson}}, \bibinfo {author} {\bibfnamefont {B.~I.}\ \bibnamefont
  {Halperin}}, \ and\ \bibinfo {author} {\bibfnamefont {C.~M.}\ \bibnamefont
  {Varma}},\ }\href@noop {} {\bibfield  {journal} {\bibinfo  {journal}
  {Philosophical Magazine}\ }\textbf {\bibinfo {volume} {25}},\ \bibinfo
  {pages} {1} (\bibinfo {year} {1972})}\BibitemShut {NoStop}%
\bibitem [{\citenamefont {Phillips}(1972)}]{Phillips1972}%
  \BibitemOpen
  \bibfield  {author} {\bibinfo {author} {\bibfnamefont {W.~A.}\ \bibnamefont
  {Phillips}},\ }\href {\doibase 10.1007/BF00660072} {\bibfield  {journal}
  {\bibinfo  {journal} {Journal of Low Temperature Physics}\ }\textbf {\bibinfo
  {volume} {7}},\ \bibinfo {pages} {351} (\bibinfo {year} {1972})}\BibitemShut
  {NoStop}%
\bibitem [{\citenamefont {Oliver}\ and\ \citenamefont
  {Welander}(2013)}]{mrsbulletin_2013}%
  \BibitemOpen
  \bibfield  {author} {\bibinfo {author} {\bibfnamefont {W.~D.}\ \bibnamefont
  {Oliver}}\ and\ \bibinfo {author} {\bibfnamefont {P.}~\bibnamefont
  {Welander}},\ }\href {\doibase 10.1557/mrs.2013.296} {\bibfield  {journal}
  {\bibinfo  {journal} {MRS Bulletin}\ }\textbf {\bibinfo {volume} {38}},\
  \bibinfo {pages} {1004–1004} (\bibinfo {year} {2013})}\BibitemShut
  {NoStop}%
\bibitem [{\citenamefont {Fowler}\ \emph {et~al.}(2012)\citenamefont {Fowler},
  \citenamefont {Mariantoni}, \citenamefont {Martinis},\ and\ \citenamefont
  {Cleland}}]{Fowler_2012}%
  \BibitemOpen
  \bibfield  {author} {\bibinfo {author} {\bibfnamefont {A.~G.}\ \bibnamefont
  {Fowler}}, \bibinfo {author} {\bibfnamefont {M.}~\bibnamefont {Mariantoni}},
  \bibinfo {author} {\bibfnamefont {J.~M.}\ \bibnamefont {Martinis}}, \ and\
  \bibinfo {author} {\bibfnamefont {A.~N.}\ \bibnamefont {Cleland}},\ }\href
  {\doibase 10.1103/physreva.86.032324} {\bibfield  {journal} {\bibinfo
  {journal} {Physical Review A}\ }\textbf {\bibinfo {volume} {86}},\ \bibinfo
  {pages} {032324} (\bibinfo {year} {2012})}\BibitemShut {NoStop}%
\bibitem [{\citenamefont {Yoder}\ \emph {et~al.}()\citenamefont {Yoder},
  \citenamefont {Kim},\ and\ \citenamefont {et~al}}]{Yoder}%
  \BibitemOpen
  \bibfield  {author} {\bibinfo {author} {\bibfnamefont {J.~L.}\ \bibnamefont
  {Yoder}}, \bibinfo {author} {\bibfnamefont {D.}~\bibnamefont {Kim}}, \ and\
  \bibinfo {author} {\bibnamefont {et~al}},\ }\href@noop {} {}\bibinfo {note}
  {To be submitted}\BibitemShut {NoStop}%
\bibitem [{\citenamefont {Chen}\ \emph {et~al.}(2014)\citenamefont {Chen},
  \citenamefont {Megrant}, \citenamefont {Kelly}, \citenamefont {Barends},
  \citenamefont {Bochmann}, \citenamefont {Chen}, \citenamefont {Chiaro},
  \citenamefont {Dunsworth}, \citenamefont {Jeffrey}, \citenamefont {Mutus},
  \citenamefont {O'Malley}, \citenamefont {Neill}, \citenamefont {Roushan},
  \citenamefont {Sank}, \citenamefont {Vainsencher}, \citenamefont {Wenner},
  \citenamefont {White}, \citenamefont {Cleland},\ and\ \citenamefont
  {Martinis}}]{Chen_2014}%
  \BibitemOpen
  \bibfield  {author} {\bibinfo {author} {\bibfnamefont {Z.}~\bibnamefont
  {Chen}}, \bibinfo {author} {\bibfnamefont {A.}~\bibnamefont {Megrant}},
  \bibinfo {author} {\bibfnamefont {J.}~\bibnamefont {Kelly}}, \bibinfo
  {author} {\bibfnamefont {R.}~\bibnamefont {Barends}}, \bibinfo {author}
  {\bibfnamefont {J.}~\bibnamefont {Bochmann}}, \bibinfo {author}
  {\bibfnamefont {Y.}~\bibnamefont {Chen}}, \bibinfo {author} {\bibfnamefont
  {B.}~\bibnamefont {Chiaro}}, \bibinfo {author} {\bibfnamefont
  {A.}~\bibnamefont {Dunsworth}}, \bibinfo {author} {\bibfnamefont
  {E.}~\bibnamefont {Jeffrey}}, \bibinfo {author} {\bibfnamefont {J.~Y.}\
  \bibnamefont {Mutus}}, \bibinfo {author} {\bibfnamefont {P.~J.~J.}\
  \bibnamefont {O'Malley}}, \bibinfo {author} {\bibfnamefont {C.}~\bibnamefont
  {Neill}}, \bibinfo {author} {\bibfnamefont {P.}~\bibnamefont {Roushan}},
  \bibinfo {author} {\bibfnamefont {D.}~\bibnamefont {Sank}}, \bibinfo {author}
  {\bibfnamefont {A.}~\bibnamefont {Vainsencher}}, \bibinfo {author}
  {\bibfnamefont {J.}~\bibnamefont {Wenner}}, \bibinfo {author} {\bibfnamefont
  {T.~C.}\ \bibnamefont {White}}, \bibinfo {author} {\bibfnamefont {A.~N.}\
  \bibnamefont {Cleland}}, \ and\ \bibinfo {author} {\bibfnamefont {J.~M.}\
  \bibnamefont {Martinis}},\ }\href {\doibase 10.1063/1.4863745} {\bibfield
  {journal} {\bibinfo  {journal} {Applied Physics Letters}\ }\textbf {\bibinfo
  {volume} {104}},\ \bibinfo {pages} {052602} (\bibinfo {year} {2014})},\
  \Eprint {http://arxiv.org/abs/https://doi.org/10.1063/1.4863745}
  {https://doi.org/10.1063/1.4863745} \BibitemShut {NoStop}%
\bibitem [{\citenamefont {Rosenberg}\ \emph {et~al.}(2016)\citenamefont
  {Rosenberg}, \citenamefont {Kim}, \citenamefont {Das}, \citenamefont {Yost},
  \citenamefont {Gustavsson}, \citenamefont {Hover}, \citenamefont {Krantz},
  \citenamefont {Melville}, \citenamefont {Racz}, \citenamefont {Samach},
  \citenamefont {Weber}, \citenamefont {Yan}, \citenamefont {Yoder},
  \citenamefont {Kerman},\ and\ \citenamefont {Oliver}}]{Rosenberg_2017}%
  \BibitemOpen
  \bibfield  {author} {\bibinfo {author} {\bibfnamefont {D.}~\bibnamefont
  {Rosenberg}}, \bibinfo {author} {\bibfnamefont {D.}~\bibnamefont {Kim}},
  \bibinfo {author} {\bibfnamefont {R.}~\bibnamefont {Das}}, \bibinfo {author}
  {\bibfnamefont {D.}~\bibnamefont {Yost}}, \bibinfo {author} {\bibfnamefont
  {S.}~\bibnamefont {Gustavsson}}, \bibinfo {author} {\bibfnamefont
  {D.}~\bibnamefont {Hover}}, \bibinfo {author} {\bibfnamefont
  {P.}~\bibnamefont {Krantz}}, \bibinfo {author} {\bibfnamefont
  {A.}~\bibnamefont {Melville}}, \bibinfo {author} {\bibfnamefont
  {L.}~\bibnamefont {Racz}}, \bibinfo {author} {\bibfnamefont {G.~O.}\
  \bibnamefont {Samach}}, \bibinfo {author} {\bibfnamefont {S.~J.}\
  \bibnamefont {Weber}}, \bibinfo {author} {\bibfnamefont {F.}~\bibnamefont
  {Yan}}, \bibinfo {author} {\bibfnamefont {J.~L.}\ \bibnamefont {Yoder}},
  \bibinfo {author} {\bibfnamefont {A.~J.}\ \bibnamefont {Kerman}}, \ and\
  \bibinfo {author} {\bibfnamefont {W.~D.}\ \bibnamefont {Oliver}},\
  }\href@noop {} {\bibfield  {journal} {\bibinfo  {journal} {npj Quantum
  Information}\ }\textbf {\bibinfo {volume} {3}},\ \bibinfo {pages} {42}
  (\bibinfo {year} {2016})}\BibitemShut {NoStop}%
\bibitem [{\citenamefont {Foxen}\ \emph
  {et~al.}(2018{\natexlab{a}})\citenamefont {Foxen}, \citenamefont {Mutus},
  \citenamefont {Lucero}, \citenamefont {Graff}, \citenamefont {Megrant},
  \citenamefont {Chen}, \citenamefont {Quintana}, \citenamefont {Burkett},
  \citenamefont {Kelly}, \citenamefont {Jeffrey}, \citenamefont {Yang},
  \citenamefont {Yu}, \citenamefont {Arya}, \citenamefont {Barends},
  \citenamefont {Chen}, \citenamefont {Chiaro}, \citenamefont {Dunsworth},
  \citenamefont {Fowler}, \citenamefont {Gidney}, \citenamefont {Giustina},
  \citenamefont {Huang}, \citenamefont {Klimov}, \citenamefont {Neeley},
  \citenamefont {Neill}, \citenamefont {Roushan}, \citenamefont {Sank},
  \citenamefont {Vainsencher}, \citenamefont {Wenner}, \citenamefont {White},\
  and\ \citenamefont {Martinis}}]{Foxen_2018}%
  \BibitemOpen
  \bibfield  {author} {\bibinfo {author} {\bibfnamefont {B.}~\bibnamefont
  {Foxen}}, \bibinfo {author} {\bibfnamefont {J.~Y.}\ \bibnamefont {Mutus}},
  \bibinfo {author} {\bibfnamefont {E.}~\bibnamefont {Lucero}}, \bibinfo
  {author} {\bibfnamefont {R.}~\bibnamefont {Graff}}, \bibinfo {author}
  {\bibfnamefont {A.}~\bibnamefont {Megrant}}, \bibinfo {author} {\bibfnamefont
  {Y.}~\bibnamefont {Chen}}, \bibinfo {author} {\bibfnamefont {C.}~\bibnamefont
  {Quintana}}, \bibinfo {author} {\bibfnamefont {B.}~\bibnamefont {Burkett}},
  \bibinfo {author} {\bibfnamefont {J.}~\bibnamefont {Kelly}}, \bibinfo
  {author} {\bibfnamefont {E.}~\bibnamefont {Jeffrey}}, \bibinfo {author}
  {\bibfnamefont {Y.}~\bibnamefont {Yang}}, \bibinfo {author} {\bibfnamefont
  {A.}~\bibnamefont {Yu}}, \bibinfo {author} {\bibfnamefont {K.}~\bibnamefont
  {Arya}}, \bibinfo {author} {\bibfnamefont {R.}~\bibnamefont {Barends}},
  \bibinfo {author} {\bibfnamefont {Z.}~\bibnamefont {Chen}}, \bibinfo {author}
  {\bibfnamefont {B.}~\bibnamefont {Chiaro}}, \bibinfo {author} {\bibfnamefont
  {A.}~\bibnamefont {Dunsworth}}, \bibinfo {author} {\bibfnamefont
  {A.}~\bibnamefont {Fowler}}, \bibinfo {author} {\bibfnamefont
  {C.}~\bibnamefont {Gidney}}, \bibinfo {author} {\bibfnamefont
  {M.}~\bibnamefont {Giustina}}, \bibinfo {author} {\bibfnamefont
  {T.}~\bibnamefont {Huang}}, \bibinfo {author} {\bibfnamefont
  {P.}~\bibnamefont {Klimov}}, \bibinfo {author} {\bibfnamefont
  {M.}~\bibnamefont {Neeley}}, \bibinfo {author} {\bibfnamefont
  {C.}~\bibnamefont {Neill}}, \bibinfo {author} {\bibfnamefont
  {P.}~\bibnamefont {Roushan}}, \bibinfo {author} {\bibfnamefont
  {D.}~\bibnamefont {Sank}}, \bibinfo {author} {\bibfnamefont {A.}~\bibnamefont
  {Vainsencher}}, \bibinfo {author} {\bibfnamefont {J.}~\bibnamefont {Wenner}},
  \bibinfo {author} {\bibfnamefont {T.~C.}\ \bibnamefont {White}}, \ and\
  \bibinfo {author} {\bibfnamefont {J.~M.}\ \bibnamefont {Martinis}},\ }\href
  {http://stacks.iop.org/2058-9565/3/i=1/a=014005} {\bibfield  {journal}
  {\bibinfo  {journal} {Quantum Science and Technology}\ }\textbf {\bibinfo
  {volume} {3}},\ \bibinfo {pages} {014005} (\bibinfo {year}
  {2018}{\natexlab{a}})}\BibitemShut {NoStop}%
\bibitem [{\citenamefont {O'Brien}\ \emph {et~al.}(2017)\citenamefont
  {O'Brien}, \citenamefont {Vahidpour}, \citenamefont {Whyland}, \citenamefont
  {Angeles}, \citenamefont {Marshall}, \citenamefont {Scarabelli},
  \citenamefont {Crossman}, \citenamefont {Yadav}, \citenamefont {Mohan},
  \citenamefont {Bui}, \citenamefont {Rawat}, \citenamefont {Renzas},
  \citenamefont {Vodrahalli}, \citenamefont {Bestwick},\ and\ \citenamefont
  {Rigetti}}]{Obrien_2017}%
  \BibitemOpen
  \bibfield  {author} {\bibinfo {author} {\bibfnamefont {W.}~\bibnamefont
  {O'Brien}}, \bibinfo {author} {\bibfnamefont {M.}~\bibnamefont {Vahidpour}},
  \bibinfo {author} {\bibfnamefont {J.~T.}\ \bibnamefont {Whyland}}, \bibinfo
  {author} {\bibfnamefont {J.}~\bibnamefont {Angeles}}, \bibinfo {author}
  {\bibfnamefont {J.}~\bibnamefont {Marshall}}, \bibinfo {author}
  {\bibfnamefont {D.}~\bibnamefont {Scarabelli}}, \bibinfo {author}
  {\bibfnamefont {G.}~\bibnamefont {Crossman}}, \bibinfo {author}
  {\bibfnamefont {K.}~\bibnamefont {Yadav}}, \bibinfo {author} {\bibfnamefont
  {Y.}~\bibnamefont {Mohan}}, \bibinfo {author} {\bibfnamefont
  {C.}~\bibnamefont {Bui}}, \bibinfo {author} {\bibfnamefont {V.}~\bibnamefont
  {Rawat}}, \bibinfo {author} {\bibfnamefont {R.}~\bibnamefont {Renzas}},
  \bibinfo {author} {\bibfnamefont {N.}~\bibnamefont {Vodrahalli}}, \bibinfo
  {author} {\bibfnamefont {A.}~\bibnamefont {Bestwick}}, \ and\ \bibinfo
  {author} {\bibfnamefont {C.}~\bibnamefont {Rigetti}},\ }\href@noop {}
  {\enquote {\bibinfo {title} {Superconducting caps for quantum integrated
  circuits},}\ } (\bibinfo {year} {2017}),\ \Eprint
  {http://arxiv.org/abs/1708.02219} {arXiv:1708.02219 [quant-ph]} \BibitemShut
  {NoStop}%
\bibitem [{\citenamefont {{Yohannes}}\ \emph {et~al.}(2009)\citenamefont
  {{Yohannes}}, \citenamefont {{Inamdar}},\ and\ \citenamefont
  {{Tolpygo}}}]{hypres}%
  \BibitemOpen
  \bibfield  {author} {\bibinfo {author} {\bibfnamefont {D.~T.}\ \bibnamefont
  {{Yohannes}}}, \bibinfo {author} {\bibfnamefont {A.}~\bibnamefont
  {{Inamdar}}}, \ and\ \bibinfo {author} {\bibfnamefont {S.~K.}\ \bibnamefont
  {{Tolpygo}}},\ }\href {\doibase 10.1109/TASC.2009.2019195} {\bibfield
  {journal} {\bibinfo  {journal} {IEEE Transactions on Applied
  Superconductivity}\ }\textbf {\bibinfo {volume} {19}},\ \bibinfo {pages}
  {149} (\bibinfo {year} {2009})}\BibitemShut {NoStop}%
\bibitem [{\citenamefont {Tolpygo}\ \emph {et~al.}(2015)\citenamefont
  {Tolpygo}, \citenamefont {Bolkhovsky}, \citenamefont {Weir}, \citenamefont
  {Johnson}, \citenamefont {Gouker},\ and\ \citenamefont
  {Oliver}}]{Tolpygo_2015}%
  \BibitemOpen
  \bibfield  {author} {\bibinfo {author} {\bibfnamefont {S.~K.}\ \bibnamefont
  {Tolpygo}}, \bibinfo {author} {\bibfnamefont {V.}~\bibnamefont {Bolkhovsky}},
  \bibinfo {author} {\bibfnamefont {T.~J.}\ \bibnamefont {Weir}}, \bibinfo
  {author} {\bibfnamefont {L.~M.}\ \bibnamefont {Johnson}}, \bibinfo {author}
  {\bibfnamefont {M.~A.}\ \bibnamefont {Gouker}}, \ and\ \bibinfo {author}
  {\bibfnamefont {W.~D.}\ \bibnamefont {Oliver}},\ }\href {\doibase
  10.1109/tasc.2014.2374836} {\bibfield  {journal} {\bibinfo  {journal} {{IEEE}
  Transactions on Applied Superconductivity}\ }\textbf {\bibinfo {volume}
  {25}},\ \bibinfo {pages} {1} (\bibinfo {year} {2015})}\BibitemShut {NoStop}%
\bibitem [{\citenamefont {Dunsworth}\ \emph {et~al.}(2018)\citenamefont
  {Dunsworth}, \citenamefont {Barends}, \citenamefont {Chen}, \citenamefont
  {Chen}, \citenamefont {Chiaro}, \citenamefont {Fowler}, \citenamefont
  {Foxen}, \citenamefont {Jeffrey}, \citenamefont {Kelly}, \citenamefont
  {Klimov}, \citenamefont {Lucero}, \citenamefont {Mutus}, \citenamefont
  {Neeley}, \citenamefont {Neill}, \citenamefont {Quintana}, \citenamefont
  {Roushan}, \citenamefont {Sank}, \citenamefont {Vainsencher}, \citenamefont
  {Wenner}, \citenamefont {White}, \citenamefont {Neven}, \citenamefont
  {Martinis},\ and\ \citenamefont {Megrant}}]{Dunsworth_2018}%
  \BibitemOpen
  \bibfield  {author} {\bibinfo {author} {\bibfnamefont {A.}~\bibnamefont
  {Dunsworth}}, \bibinfo {author} {\bibfnamefont {R.}~\bibnamefont {Barends}},
  \bibinfo {author} {\bibfnamefont {Y.}~\bibnamefont {Chen}}, \bibinfo {author}
  {\bibfnamefont {Z.}~\bibnamefont {Chen}}, \bibinfo {author} {\bibfnamefont
  {B.}~\bibnamefont {Chiaro}}, \bibinfo {author} {\bibfnamefont
  {A.}~\bibnamefont {Fowler}}, \bibinfo {author} {\bibfnamefont
  {B.}~\bibnamefont {Foxen}}, \bibinfo {author} {\bibfnamefont
  {E.}~\bibnamefont {Jeffrey}}, \bibinfo {author} {\bibfnamefont
  {J.}~\bibnamefont {Kelly}}, \bibinfo {author} {\bibfnamefont {P.~V.}\
  \bibnamefont {Klimov}}, \bibinfo {author} {\bibfnamefont {E.}~\bibnamefont
  {Lucero}}, \bibinfo {author} {\bibfnamefont {J.~Y.}\ \bibnamefont {Mutus}},
  \bibinfo {author} {\bibfnamefont {M.}~\bibnamefont {Neeley}}, \bibinfo
  {author} {\bibfnamefont {C.}~\bibnamefont {Neill}}, \bibinfo {author}
  {\bibfnamefont {C.}~\bibnamefont {Quintana}}, \bibinfo {author}
  {\bibfnamefont {P.}~\bibnamefont {Roushan}}, \bibinfo {author} {\bibfnamefont
  {D.}~\bibnamefont {Sank}}, \bibinfo {author} {\bibfnamefont {A.}~\bibnamefont
  {Vainsencher}}, \bibinfo {author} {\bibfnamefont {J.}~\bibnamefont {Wenner}},
  \bibinfo {author} {\bibfnamefont {T.~C.}\ \bibnamefont {White}}, \bibinfo
  {author} {\bibfnamefont {H.}~\bibnamefont {Neven}}, \bibinfo {author}
  {\bibfnamefont {J.~M.}\ \bibnamefont {Martinis}}, \ and\ \bibinfo {author}
  {\bibfnamefont {A.}~\bibnamefont {Megrant}},\ }\href {\doibase
  10.1063/1.5014033} {\bibfield  {journal} {\bibinfo  {journal} {Applied
  Physics Letters}\ }\textbf {\bibinfo {volume} {112}},\ \bibinfo {pages}
  {063502} (\bibinfo {year} {2018})},\ \Eprint
  {http://arxiv.org/abs/https://doi.org/10.1063/1.5014033}
  {https://doi.org/10.1063/1.5014033} \BibitemShut {NoStop}%
\bibitem [{\citenamefont {Vahidpour}\ \emph {et~al.}(2017)\citenamefont
  {Vahidpour}, \citenamefont {O'Brien}, \citenamefont {Whyland}, \citenamefont
  {Angeles}, \citenamefont {Marshall}, \citenamefont {Scarabelli},
  \citenamefont {Crossman}, \citenamefont {Yadav}, \citenamefont {Mohan},
  \citenamefont {Bui}, \citenamefont {Rawat}, \citenamefont {Renzas},
  \citenamefont {Vodrahalli}, \citenamefont {Bestwick},\ and\ \citenamefont
  {Rigetti}}]{Vahidpour_2017}%
  \BibitemOpen
  \bibfield  {author} {\bibinfo {author} {\bibfnamefont {M.}~\bibnamefont
  {Vahidpour}}, \bibinfo {author} {\bibfnamefont {W.}~\bibnamefont {O'Brien}},
  \bibinfo {author} {\bibfnamefont {J.~T.}\ \bibnamefont {Whyland}}, \bibinfo
  {author} {\bibfnamefont {J.}~\bibnamefont {Angeles}}, \bibinfo {author}
  {\bibfnamefont {J.}~\bibnamefont {Marshall}}, \bibinfo {author}
  {\bibfnamefont {D.}~\bibnamefont {Scarabelli}}, \bibinfo {author}
  {\bibfnamefont {G.}~\bibnamefont {Crossman}}, \bibinfo {author}
  {\bibfnamefont {K.}~\bibnamefont {Yadav}}, \bibinfo {author} {\bibfnamefont
  {Y.}~\bibnamefont {Mohan}}, \bibinfo {author} {\bibfnamefont
  {C.}~\bibnamefont {Bui}}, \bibinfo {author} {\bibfnamefont {V.}~\bibnamefont
  {Rawat}}, \bibinfo {author} {\bibfnamefont {R.}~\bibnamefont {Renzas}},
  \bibinfo {author} {\bibfnamefont {N.}~\bibnamefont {Vodrahalli}}, \bibinfo
  {author} {\bibfnamefont {A.}~\bibnamefont {Bestwick}}, \ and\ \bibinfo
  {author} {\bibfnamefont {C.}~\bibnamefont {Rigetti}},\ }\href@noop {}
  {\enquote {\bibinfo {title} {Superconducting through-silicon vias for quantum
  integrated circuits},}\ } (\bibinfo {year} {2017}),\ \Eprint
  {http://arxiv.org/abs/1708.02226} {arXiv:1708.02226 [quant-ph]} \BibitemShut
  {NoStop}%
\bibitem [{\citenamefont {Versluis}\ \emph {et~al.}(2016)\citenamefont
  {Versluis}, \citenamefont {Poletto}, \citenamefont {Khammassi}, \citenamefont
  {Haider}, \citenamefont {Michalak}, \citenamefont {Bruno}, \citenamefont
  {Bertels},\ and\ \citenamefont {DiCarlo}}]{Versluis_2016}%
  \BibitemOpen
  \bibfield  {author} {\bibinfo {author} {\bibfnamefont {R.}~\bibnamefont
  {Versluis}}, \bibinfo {author} {\bibfnamefont {S.}~\bibnamefont {Poletto}},
  \bibinfo {author} {\bibfnamefont {N.}~\bibnamefont {Khammassi}}, \bibinfo
  {author} {\bibfnamefont {N.}~\bibnamefont {Haider}}, \bibinfo {author}
  {\bibfnamefont {D.~J.}\ \bibnamefont {Michalak}}, \bibinfo {author}
  {\bibfnamefont {A.}~\bibnamefont {Bruno}}, \bibinfo {author} {\bibfnamefont
  {K.}~\bibnamefont {Bertels}}, \ and\ \bibinfo {author} {\bibfnamefont
  {L.}~\bibnamefont {DiCarlo}},\ }\href@noop {} {\enquote {\bibinfo {title}
  {Scalable quantum circuit and control for a superconducting surface code},}\
  } (\bibinfo {year} {2016}),\ \Eprint {http://arxiv.org/abs/1612.08208}
  {arXiv:1612.08208 [quant-ph]} \BibitemShut {NoStop}%
\bibitem [{\citenamefont {Mallek}\ \emph {et~al.}()\citenamefont {Mallek},
  \citenamefont {Yost},\ and\ \citenamefont {et~al}}]{Mallek}%
  \BibitemOpen
  \bibfield  {author} {\bibinfo {author} {\bibfnamefont {J.}~\bibnamefont
  {Mallek}}, \bibinfo {author} {\bibfnamefont {D.~R.}\ \bibnamefont {Yost}}, \
  and\ \bibinfo {author} {\bibnamefont {et~al}},\ }\href@noop {} {}\bibinfo
  {note} {To be submitted}\BibitemShut {NoStop}%
\bibitem [{\citenamefont {Martinis}\ \emph {et~al.}(2005)\citenamefont
  {Martinis}, \citenamefont {Cooper}, \citenamefont {McDermott}, \citenamefont
  {Steffen}, \citenamefont {Ansmann}, \citenamefont {Osborn}, \citenamefont
  {Cicak}, \citenamefont {Oh}, \citenamefont {Pappas}, \citenamefont
  {Simmonds},\ and\ \citenamefont {Yu}}]{Martinis_2005}%
  \BibitemOpen
  \bibfield  {author} {\bibinfo {author} {\bibfnamefont {J.~M.}\ \bibnamefont
  {Martinis}}, \bibinfo {author} {\bibfnamefont {K.~B.}\ \bibnamefont
  {Cooper}}, \bibinfo {author} {\bibfnamefont {R.}~\bibnamefont {McDermott}},
  \bibinfo {author} {\bibfnamefont {M.}~\bibnamefont {Steffen}}, \bibinfo
  {author} {\bibfnamefont {M.}~\bibnamefont {Ansmann}}, \bibinfo {author}
  {\bibfnamefont {K.~D.}\ \bibnamefont {Osborn}}, \bibinfo {author}
  {\bibfnamefont {K.}~\bibnamefont {Cicak}}, \bibinfo {author} {\bibfnamefont
  {S.}~\bibnamefont {Oh}}, \bibinfo {author} {\bibfnamefont {D.~P.}\
  \bibnamefont {Pappas}}, \bibinfo {author} {\bibfnamefont {R.~W.}\
  \bibnamefont {Simmonds}}, \ and\ \bibinfo {author} {\bibfnamefont {C.~C.}\
  \bibnamefont {Yu}},\ }\href {\doibase 10.1103/PhysRevLett.95.210503}
  {\bibfield  {journal} {\bibinfo  {journal} {Phys. Rev. Lett.}\ }\textbf
  {\bibinfo {volume} {95}},\ \bibinfo {pages} {210503} (\bibinfo {year}
  {2005})}\BibitemShut {NoStop}%
\bibitem [{\citenamefont {Yost}\ \emph {et~al.}()\citenamefont {Yost},
  \citenamefont {Mallek},\ and\ \citenamefont {et~al}}]{Yost}%
  \BibitemOpen
  \bibfield  {author} {\bibinfo {author} {\bibfnamefont {D.~R.}\ \bibnamefont
  {Yost}}, \bibinfo {author} {\bibfnamefont {J.}~\bibnamefont {Mallek}}, \ and\
  \bibinfo {author} {\bibnamefont {et~al}},\ }\href@noop {} {}\bibinfo {note}
  {To be submitted}\BibitemShut {NoStop}%
\bibitem [{\citenamefont {Wenner}\ \emph {et~al.}(2011)\citenamefont {Wenner},
  \citenamefont {Neeley}, \citenamefont {Bialczak}, \citenamefont {Lenander},
  \citenamefont {Lucero}, \citenamefont {O'Connell}, \citenamefont {Sank},
  \citenamefont {Wang}, \citenamefont {Weides}, \citenamefont {Cleland},\ and\
  \citenamefont {Martinis}}]{Wenner_2011}%
  \BibitemOpen
  \bibfield  {author} {\bibinfo {author} {\bibfnamefont {J.}~\bibnamefont
  {Wenner}}, \bibinfo {author} {\bibfnamefont {M.}~\bibnamefont {Neeley}},
  \bibinfo {author} {\bibfnamefont {R.~C.}\ \bibnamefont {Bialczak}}, \bibinfo
  {author} {\bibfnamefont {M.}~\bibnamefont {Lenander}}, \bibinfo {author}
  {\bibfnamefont {E.}~\bibnamefont {Lucero}}, \bibinfo {author} {\bibfnamefont
  {A.~D.}\ \bibnamefont {O'Connell}}, \bibinfo {author} {\bibfnamefont
  {D.}~\bibnamefont {Sank}}, \bibinfo {author} {\bibfnamefont {H.}~\bibnamefont
  {Wang}}, \bibinfo {author} {\bibfnamefont {M.}~\bibnamefont {Weides}},
  \bibinfo {author} {\bibfnamefont {A.~N.}\ \bibnamefont {Cleland}}, \ and\
  \bibinfo {author} {\bibfnamefont {J.~M.}\ \bibnamefont {Martinis}},\ }\href
  {\doibase 10.1088/0953-2048/24/6/065001} {\bibfield  {journal} {\bibinfo
  {journal} {Superconductor Science and Technology}\ }\textbf {\bibinfo
  {volume} {24}},\ \bibinfo {pages} {065001} (\bibinfo {year}
  {2011})}\BibitemShut {NoStop}%
\bibitem [{\citenamefont {Brecht}\ \emph {et~al.}(2016)\citenamefont {Brecht},
  \citenamefont {Pfaff}, \citenamefont {Wang}, \citenamefont {Chu},
  \citenamefont {Frunzio}, \citenamefont {Devoret},\ and\ \citenamefont
  {Schoelkopf}}]{Brecht_2016}%
  \BibitemOpen
  \bibfield  {author} {\bibinfo {author} {\bibfnamefont {T.}~\bibnamefont
  {Brecht}}, \bibinfo {author} {\bibfnamefont {W.}~\bibnamefont {Pfaff}},
  \bibinfo {author} {\bibfnamefont {C.}~\bibnamefont {Wang}}, \bibinfo {author}
  {\bibfnamefont {Y.}~\bibnamefont {Chu}}, \bibinfo {author} {\bibfnamefont
  {L.}~\bibnamefont {Frunzio}}, \bibinfo {author} {\bibfnamefont {M.~H.}\
  \bibnamefont {Devoret}}, \ and\ \bibinfo {author} {\bibfnamefont {R.~J.}\
  \bibnamefont {Schoelkopf}},\ }\href@noop {} {\bibfield  {journal} {\bibinfo
  {journal} {npj Quantum Information}\ }\textbf {\bibinfo {volume} {2}},\
  \bibinfo {pages} {16002} (\bibinfo {year} {2016})}\BibitemShut {NoStop}%
\bibitem [{\citenamefont {Nigg}\ \emph {et~al.}(2012)\citenamefont {Nigg},
  \citenamefont {Paik}, \citenamefont {Vlastakis}, \citenamefont {Kirchmair},
  \citenamefont {Shankar}, \citenamefont {Frunzio}, \citenamefont {Devoret},
  \citenamefont {Schoelkopf},\ and\ \citenamefont {Girvin}}]{BBQ}%
  \BibitemOpen
  \bibfield  {author} {\bibinfo {author} {\bibfnamefont {S.~E.}\ \bibnamefont
  {Nigg}}, \bibinfo {author} {\bibfnamefont {H.}~\bibnamefont {Paik}}, \bibinfo
  {author} {\bibfnamefont {B.}~\bibnamefont {Vlastakis}}, \bibinfo {author}
  {\bibfnamefont {G.}~\bibnamefont {Kirchmair}}, \bibinfo {author}
  {\bibfnamefont {S.}~\bibnamefont {Shankar}}, \bibinfo {author} {\bibfnamefont
  {L.}~\bibnamefont {Frunzio}}, \bibinfo {author} {\bibfnamefont {M.~H.}\
  \bibnamefont {Devoret}}, \bibinfo {author} {\bibfnamefont {R.~J.}\
  \bibnamefont {Schoelkopf}}, \ and\ \bibinfo {author} {\bibfnamefont {S.~M.}\
  \bibnamefont {Girvin}},\ }\href {\doibase 10.1103/PhysRevLett.108.240502}
  {\bibfield  {journal} {\bibinfo  {journal} {Phys. Rev. Lett.}\ }\textbf
  {\bibinfo {volume} {108}},\ \bibinfo {pages} {240502} (\bibinfo {year}
  {2012})}\BibitemShut {NoStop}%
\bibitem [{\citenamefont {Solgun}\ \emph {et~al.}(2014)\citenamefont {Solgun},
  \citenamefont {Abraham},\ and\ \citenamefont {DiVincenzo}}]{Brune}%
  \BibitemOpen
  \bibfield  {author} {\bibinfo {author} {\bibfnamefont {F.}~\bibnamefont
  {Solgun}}, \bibinfo {author} {\bibfnamefont {D.~W.}\ \bibnamefont {Abraham}},
  \ and\ \bibinfo {author} {\bibfnamefont {D.~P.}\ \bibnamefont {DiVincenzo}},\
  }\href {\doibase 10.1103/PhysRevB.90.134504} {\bibfield  {journal} {\bibinfo
  {journal} {Phys. Rev. B}\ }\textbf {\bibinfo {volume} {90}},\ \bibinfo
  {pages} {134504} (\bibinfo {year} {2014})}\BibitemShut {NoStop}%
\bibitem [{\citenamefont {Brecht}\ \emph {et~al.}(2017)\citenamefont {Brecht},
  \citenamefont {Chu}, \citenamefont {Axline}, \citenamefont {Pfaff},
  \citenamefont {Blumoff}, \citenamefont {Chou}, \citenamefont {Krayzman},
  \citenamefont {Frunzio},\ and\ \citenamefont {Schoelkopf}}]{Brecht_2017}%
  \BibitemOpen
  \bibfield  {author} {\bibinfo {author} {\bibfnamefont {T.}~\bibnamefont
  {Brecht}}, \bibinfo {author} {\bibfnamefont {Y.}~\bibnamefont {Chu}},
  \bibinfo {author} {\bibfnamefont {C.}~\bibnamefont {Axline}}, \bibinfo
  {author} {\bibfnamefont {W.}~\bibnamefont {Pfaff}}, \bibinfo {author}
  {\bibfnamefont {J.~Z.}\ \bibnamefont {Blumoff}}, \bibinfo {author}
  {\bibfnamefont {K.}~\bibnamefont {Chou}}, \bibinfo {author} {\bibfnamefont
  {L.}~\bibnamefont {Krayzman}}, \bibinfo {author} {\bibfnamefont
  {L.}~\bibnamefont {Frunzio}}, \ and\ \bibinfo {author} {\bibfnamefont
  {R.~J.}\ \bibnamefont {Schoelkopf}},\ }\href {\doibase
  10.1103/PhysRevApplied.7.044018} {\bibfield  {journal} {\bibinfo  {journal}
  {Phys. Rev. Applied}\ }\textbf {\bibinfo {volume} {7}},\ \bibinfo {pages}
  {044018} (\bibinfo {year} {2017})}\BibitemShut {NoStop}%
\bibitem [{\citenamefont {Colless}\ and\ \citenamefont
  {Reilly}(2014)}]{Colless_2014}%
  \BibitemOpen
  \bibfield  {author} {\bibinfo {author} {\bibfnamefont {J.~I.}\ \bibnamefont
  {Colless}}\ and\ \bibinfo {author} {\bibfnamefont {D.~J.}\ \bibnamefont
  {Reilly}},\ }\href {\doibase 10.1063/1.4900948} {\bibfield  {journal}
  {\bibinfo  {journal} {Review of Scientific Instruments}\ }\textbf {\bibinfo
  {volume} {85}},\ \bibinfo {pages} {114706} (\bibinfo {year} {2014})},\
  \Eprint {http://arxiv.org/abs/https://doi.org/10.1063/1.4900948}
  {https://doi.org/10.1063/1.4900948} \BibitemShut {NoStop}%
\bibitem [{\citenamefont {Caudillo}\ \emph {et~al.}(2018)\citenamefont
  {Caudillo}, \citenamefont {Yoscovits}, \citenamefont {Lampert}, \citenamefont
  {Michalak}, \citenamefont {Elsherbini}, \citenamefont {Falcon}, \citenamefont
  {Roberts}, \citenamefont {DiCarlo},\ and\ \citenamefont
  {Clarke}}]{Caudillo_2018}%
  \BibitemOpen
  \bibfield  {author} {\bibinfo {author} {\bibfnamefont {R.}~\bibnamefont
  {Caudillo}}, \bibinfo {author} {\bibfnamefont {Z.}~\bibnamefont {Yoscovits}},
  \bibinfo {author} {\bibfnamefont {L.}~\bibnamefont {Lampert}}, \bibinfo
  {author} {\bibfnamefont {D.}~\bibnamefont {Michalak}}, \bibinfo {author}
  {\bibfnamefont {A.}~\bibnamefont {Elsherbini}}, \bibinfo {author}
  {\bibfnamefont {J.}~\bibnamefont {Falcon}}, \bibinfo {author} {\bibfnamefont
  {J.}~\bibnamefont {Roberts}}, \bibinfo {author} {\bibfnamefont
  {L.}~\bibnamefont {DiCarlo}}, \ and\ \bibinfo {author} {\bibfnamefont
  {J.}~\bibnamefont {Clarke}},\ }\href@noop {} {\enquote {\bibinfo {title} {Die
  design and fabrication for flip-chip-packaged superconducting quantum
  processors},}\ } (\bibinfo {year} {2018}),\ \bibinfo {note} {aPS March
  Meeting 2018}\BibitemShut {NoStop}%
\bibitem [{\citenamefont {Bronn}\ \emph {et~al.}(2018)\citenamefont {Bronn},
  \citenamefont {Adiga}, \citenamefont {Olivadese}, \citenamefont {Wu},
  \citenamefont {Chow},\ and\ \citenamefont {Pappas}}]{Bronn_2018}%
  \BibitemOpen
  \bibfield  {author} {\bibinfo {author} {\bibfnamefont {N.~T.}\ \bibnamefont
  {Bronn}}, \bibinfo {author} {\bibfnamefont {V.~P.}\ \bibnamefont {Adiga}},
  \bibinfo {author} {\bibfnamefont {S.~B.}\ \bibnamefont {Olivadese}}, \bibinfo
  {author} {\bibfnamefont {X.}~\bibnamefont {Wu}}, \bibinfo {author}
  {\bibfnamefont {J.~M.}\ \bibnamefont {Chow}}, \ and\ \bibinfo {author}
  {\bibfnamefont {D.~P.}\ \bibnamefont {Pappas}},\ }\href
  {http://stacks.iop.org/2058-9565/3/i=2/a=024007} {\bibfield  {journal}
  {\bibinfo  {journal} {Quantum Science and Technology}\ }\textbf {\bibinfo
  {volume} {3}},\ \bibinfo {pages} {024007} (\bibinfo {year}
  {2018})}\BibitemShut {NoStop}%
\bibitem [{\citenamefont {Rahamim}\ \emph {et~al.}(2017)\citenamefont
  {Rahamim}, \citenamefont {Behrle}, \citenamefont {Peterer}, \citenamefont
  {Patterson}, \citenamefont {Spring}, \citenamefont {Tsunoda}, \citenamefont
  {Manenti}, \citenamefont {Tancredi},\ and\ \citenamefont
  {Leek}}]{Rahanim_2017}%
  \BibitemOpen
  \bibfield  {author} {\bibinfo {author} {\bibfnamefont {J.}~\bibnamefont
  {Rahamim}}, \bibinfo {author} {\bibfnamefont {T.}~\bibnamefont {Behrle}},
  \bibinfo {author} {\bibfnamefont {M.~J.}\ \bibnamefont {Peterer}}, \bibinfo
  {author} {\bibfnamefont {A.}~\bibnamefont {Patterson}}, \bibinfo {author}
  {\bibfnamefont {P.~A.}\ \bibnamefont {Spring}}, \bibinfo {author}
  {\bibfnamefont {T.}~\bibnamefont {Tsunoda}}, \bibinfo {author} {\bibfnamefont
  {R.}~\bibnamefont {Manenti}}, \bibinfo {author} {\bibfnamefont
  {G.}~\bibnamefont {Tancredi}}, \ and\ \bibinfo {author} {\bibfnamefont
  {P.~J.}\ \bibnamefont {Leek}},\ }\href {\doibase 10.1063/1.4984299}
  {\bibfield  {journal} {\bibinfo  {journal} {Applied Physics Letters}\
  }\textbf {\bibinfo {volume} {110}},\ \bibinfo {pages} {222602} (\bibinfo
  {year} {2017})},\ \Eprint
  {http://arxiv.org/abs/https://doi.org/10.1063/1.4984299}
  {https://doi.org/10.1063/1.4984299} \BibitemShut {NoStop}%
\bibitem [{\citenamefont {Patterson}\ \emph {et~al.}(2019)\citenamefont
  {Patterson}, \citenamefont {Rahamim}, \citenamefont {Tsunoda}, \citenamefont
  {Spring}, \citenamefont {Jebari}, \citenamefont {Ratter}, \citenamefont
  {Mergenthaler}, \citenamefont {Tancredi}, \citenamefont {Vlastakis},
  \citenamefont {Esposito},\ and\ \citenamefont {Leek}}]{Patterson_2019}%
  \BibitemOpen
  \bibfield  {author} {\bibinfo {author} {\bibfnamefont {A.}~\bibnamefont
  {Patterson}}, \bibinfo {author} {\bibfnamefont {J.}~\bibnamefont {Rahamim}},
  \bibinfo {author} {\bibfnamefont {T.}~\bibnamefont {Tsunoda}}, \bibinfo
  {author} {\bibfnamefont {P.~A.}\ \bibnamefont {Spring}}, \bibinfo {author}
  {\bibfnamefont {S.}~\bibnamefont {Jebari}}, \bibinfo {author} {\bibfnamefont
  {K.}~\bibnamefont {Ratter}}, \bibinfo {author} {\bibfnamefont
  {M.}~\bibnamefont {Mergenthaler}}, \bibinfo {author} {\bibfnamefont
  {G.}~\bibnamefont {Tancredi}}, \bibinfo {author} {\bibfnamefont
  {B.}~\bibnamefont {Vlastakis}}, \bibinfo {author} {\bibfnamefont
  {M.}~\bibnamefont {Esposito}}, \ and\ \bibinfo {author} {\bibfnamefont
  {P.~J.}\ \bibnamefont {Leek}},\ }\href@noop {} {\enquote {\bibinfo {title}
  {Calibration of the cross-resonance two-qubit gate between directly-coupled
  transmons},}\ } (\bibinfo {year} {2019}),\ \Eprint
  {http://arxiv.org/abs/1905.05670} {arXiv:1905.05670 [quant-ph]} \BibitemShut
  {NoStop}%
\bibitem [{\citenamefont {Liu}\ \emph {et~al.}(2017)\citenamefont {Liu},
  \citenamefont {Li}, \citenamefont {Dai}, \citenamefont {Zhang}, \citenamefont
  {Xue}, \citenamefont {Tan}, \citenamefont {Yu},\ and\ \citenamefont
  {Yu}}]{Liu_2017}%
  \BibitemOpen
  \bibfield  {author} {\bibinfo {author} {\bibfnamefont {Q.}~\bibnamefont
  {Liu}}, \bibinfo {author} {\bibfnamefont {M.}~\bibnamefont {Li}}, \bibinfo
  {author} {\bibfnamefont {K.}~\bibnamefont {Dai}}, \bibinfo {author}
  {\bibfnamefont {K.}~\bibnamefont {Zhang}}, \bibinfo {author} {\bibfnamefont
  {G.}~\bibnamefont {Xue}}, \bibinfo {author} {\bibfnamefont {X.}~\bibnamefont
  {Tan}}, \bibinfo {author} {\bibfnamefont {H.}~\bibnamefont {Yu}}, \ and\
  \bibinfo {author} {\bibfnamefont {Y.}~\bibnamefont {Yu}},\ }\href {\doibase
  10.1063/1.4985435} {\bibfield  {journal} {\bibinfo  {journal} {Applied
  Physics Letters}\ }\textbf {\bibinfo {volume} {110}},\ \bibinfo {pages}
  {232602} (\bibinfo {year} {2017})},\ \Eprint
  {http://arxiv.org/abs/https://doi.org/10.1063/1.4985435}
  {https://doi.org/10.1063/1.4985435} \BibitemShut {NoStop}%
\bibitem [{\citenamefont {B{\'{e}}janin}\ \emph {et~al.}(2016)\citenamefont
  {B{\'{e}}janin}, \citenamefont {McConkey}, \citenamefont {Rinehart},
  \citenamefont {Earnest}, \citenamefont {McRae}, \citenamefont {Shiri},
  \citenamefont {Bateman}, \citenamefont {Rohanizadegan}, \citenamefont
  {Penava}, \citenamefont {Breul}, \citenamefont {Royak}, \citenamefont
  {Zapatka}, \citenamefont {Fowler},\ and\ \citenamefont
  {Mariantoni}}]{Bejanin_2016}%
  \BibitemOpen
  \bibfield  {author} {\bibinfo {author} {\bibfnamefont {J.}~\bibnamefont
  {B{\'{e}}janin}}, \bibinfo {author} {\bibfnamefont {T.}~\bibnamefont
  {McConkey}}, \bibinfo {author} {\bibfnamefont {J.}~\bibnamefont {Rinehart}},
  \bibinfo {author} {\bibfnamefont {C.}~\bibnamefont {Earnest}}, \bibinfo
  {author} {\bibfnamefont {C.}~\bibnamefont {McRae}}, \bibinfo {author}
  {\bibfnamefont {D.}~\bibnamefont {Shiri}}, \bibinfo {author} {\bibfnamefont
  {J.}~\bibnamefont {Bateman}}, \bibinfo {author} {\bibfnamefont
  {Y.}~\bibnamefont {Rohanizadegan}}, \bibinfo {author} {\bibfnamefont
  {B.}~\bibnamefont {Penava}}, \bibinfo {author} {\bibfnamefont
  {P.}~\bibnamefont {Breul}}, \bibinfo {author} {\bibfnamefont
  {S.}~\bibnamefont {Royak}}, \bibinfo {author} {\bibfnamefont
  {M.}~\bibnamefont {Zapatka}}, \bibinfo {author} {\bibfnamefont
  {A.}~\bibnamefont {Fowler}}, \ and\ \bibinfo {author} {\bibfnamefont
  {M.}~\bibnamefont {Mariantoni}},\ }\href {\doibase
  10.1103/physrevapplied.6.044010} {\bibfield  {journal} {\bibinfo  {journal}
  {Physical Review Applied}\ }\textbf {\bibinfo {volume} {6}},\ \bibinfo
  {pages} {044010} (\bibinfo {year} {2016})}\BibitemShut {NoStop}%
\bibitem [{\citenamefont {Foxen}\ \emph
  {et~al.}(2018{\natexlab{b}})\citenamefont {Foxen}, \citenamefont {Mutus},
  \citenamefont {Lucero}, \citenamefont {Jeffrey}, \citenamefont {Sank},
  \citenamefont {Barends}, \citenamefont {Arya}, \citenamefont {Burkett},
  \citenamefont {Chen}, \citenamefont {Chen}, \citenamefont {Chiaro},
  \citenamefont {Dunsworth}, \citenamefont {Fowler}, \citenamefont {Gidney},
  \citenamefont {Giustina}, \citenamefont {Graff}, \citenamefont {Huang},
  \citenamefont {Kelly}, \citenamefont {Klimov}, \citenamefont {Megrant},
  \citenamefont {Naaman}, \citenamefont {Neeley}, \citenamefont {Neill},
  \citenamefont {Quintana}, \citenamefont {Roushan}, \citenamefont
  {Vainsencher}, \citenamefont {Wenner}, \citenamefont {White},\ and\
  \citenamefont {Martinis}}]{Foxen_2018_arxiv}%
  \BibitemOpen
  \bibfield  {author} {\bibinfo {author} {\bibfnamefont {B.}~\bibnamefont
  {Foxen}}, \bibinfo {author} {\bibfnamefont {J.}~\bibnamefont {Mutus}},
  \bibinfo {author} {\bibfnamefont {E.}~\bibnamefont {Lucero}}, \bibinfo
  {author} {\bibfnamefont {E.}~\bibnamefont {Jeffrey}}, \bibinfo {author}
  {\bibfnamefont {D.}~\bibnamefont {Sank}}, \bibinfo {author} {\bibfnamefont
  {R.}~\bibnamefont {Barends}}, \bibinfo {author} {\bibfnamefont
  {K.}~\bibnamefont {Arya}}, \bibinfo {author} {\bibfnamefont {B.}~\bibnamefont
  {Burkett}}, \bibinfo {author} {\bibfnamefont {Y.}~\bibnamefont {Chen}},
  \bibinfo {author} {\bibfnamefont {Z.}~\bibnamefont {Chen}}, \bibinfo {author}
  {\bibfnamefont {B.}~\bibnamefont {Chiaro}}, \bibinfo {author} {\bibfnamefont
  {A.}~\bibnamefont {Dunsworth}}, \bibinfo {author} {\bibfnamefont
  {A.}~\bibnamefont {Fowler}}, \bibinfo {author} {\bibfnamefont
  {C.}~\bibnamefont {Gidney}}, \bibinfo {author} {\bibfnamefont
  {M.}~\bibnamefont {Giustina}}, \bibinfo {author} {\bibfnamefont
  {R.}~\bibnamefont {Graff}}, \bibinfo {author} {\bibfnamefont
  {T.}~\bibnamefont {Huang}}, \bibinfo {author} {\bibfnamefont
  {J.}~\bibnamefont {Kelly}}, \bibinfo {author} {\bibfnamefont
  {P.}~\bibnamefont {Klimov}}, \bibinfo {author} {\bibfnamefont
  {A.}~\bibnamefont {Megrant}}, \bibinfo {author} {\bibfnamefont
  {O.}~\bibnamefont {Naaman}}, \bibinfo {author} {\bibfnamefont
  {M.}~\bibnamefont {Neeley}}, \bibinfo {author} {\bibfnamefont
  {C.}~\bibnamefont {Neill}}, \bibinfo {author} {\bibfnamefont
  {C.}~\bibnamefont {Quintana}}, \bibinfo {author} {\bibfnamefont
  {P.}~\bibnamefont {Roushan}}, \bibinfo {author} {\bibfnamefont
  {A.}~\bibnamefont {Vainsencher}}, \bibinfo {author} {\bibfnamefont
  {J.}~\bibnamefont {Wenner}}, \bibinfo {author} {\bibfnamefont
  {T.}~\bibnamefont {White}}, \ and\ \bibinfo {author} {\bibfnamefont {J.~M.}\
  \bibnamefont {Martinis}},\ }\href@noop {} {\enquote {\bibinfo {title} {High
  speed flux sampling for tunable superconducting qubits with an embedded
  cryogenic transducer},}\ } (\bibinfo {year} {2018}{\natexlab{b}}),\ \Eprint
  {http://arxiv.org/abs/1808.09612} {arXiv:1808.09612 [quant-ph]} \BibitemShut
  {NoStop}%
\bibitem [{\citenamefont {Krinner}\ \emph {et~al.}(2018)\citenamefont
  {Krinner}, \citenamefont {Storz}, \citenamefont {Kurpiers}, \citenamefont
  {Magnard}, \citenamefont {Heinsoo}, \citenamefont {Keller}, \citenamefont
  {Luetolf}, \citenamefont {Eichler},\ and\ \citenamefont
  {Wallraff}}]{Krinner_2018}%
  \BibitemOpen
  \bibfield  {author} {\bibinfo {author} {\bibfnamefont {S.}~\bibnamefont
  {Krinner}}, \bibinfo {author} {\bibfnamefont {S.}~\bibnamefont {Storz}},
  \bibinfo {author} {\bibfnamefont {P.}~\bibnamefont {Kurpiers}}, \bibinfo
  {author} {\bibfnamefont {P.}~\bibnamefont {Magnard}}, \bibinfo {author}
  {\bibfnamefont {J.}~\bibnamefont {Heinsoo}}, \bibinfo {author} {\bibfnamefont
  {R.}~\bibnamefont {Keller}}, \bibinfo {author} {\bibfnamefont
  {J.}~\bibnamefont {Luetolf}}, \bibinfo {author} {\bibfnamefont
  {C.}~\bibnamefont {Eichler}}, \ and\ \bibinfo {author} {\bibfnamefont
  {A.}~\bibnamefont {Wallraff}},\ }\href@noop {} {\enquote {\bibinfo {title}
  {Engineering cryogenic setups for 100-qubit scale superconducting circuit
  systems},}\ } (\bibinfo {year} {2018}),\ \Eprint
  {http://arxiv.org/abs/1806.07862} {arXiv:1806.07862 [quant-ph]} \BibitemShut
  {NoStop}%
\bibitem [{\citenamefont {Barends}\ \emph {et~al.}(2011)\citenamefont
  {Barends}, \citenamefont {Wenner}, \citenamefont {Lenander}, \citenamefont
  {Chen}, \citenamefont {Bialczak}, \citenamefont {Kelly}, \citenamefont
  {Lucero}, \citenamefont {O’Malley}, \citenamefont {Mariantoni},
  \citenamefont {Sank}, \citenamefont {Wang}, \citenamefont {White},
  \citenamefont {Yin}, \citenamefont {Zhao}, \citenamefont {Cleland},
  \citenamefont {Martinis},\ and\ \citenamefont {Baselmans}}]{Barends_2011}%
  \BibitemOpen
  \bibfield  {author} {\bibinfo {author} {\bibfnamefont {R.}~\bibnamefont
  {Barends}}, \bibinfo {author} {\bibfnamefont {J.}~\bibnamefont {Wenner}},
  \bibinfo {author} {\bibfnamefont {M.}~\bibnamefont {Lenander}}, \bibinfo
  {author} {\bibfnamefont {Y.}~\bibnamefont {Chen}}, \bibinfo {author}
  {\bibfnamefont {R.~C.}\ \bibnamefont {Bialczak}}, \bibinfo {author}
  {\bibfnamefont {J.}~\bibnamefont {Kelly}}, \bibinfo {author} {\bibfnamefont
  {E.}~\bibnamefont {Lucero}}, \bibinfo {author} {\bibfnamefont
  {P.}~\bibnamefont {O’Malley}}, \bibinfo {author} {\bibfnamefont
  {M.}~\bibnamefont {Mariantoni}}, \bibinfo {author} {\bibfnamefont
  {D.}~\bibnamefont {Sank}}, \bibinfo {author} {\bibfnamefont {H.}~\bibnamefont
  {Wang}}, \bibinfo {author} {\bibfnamefont {T.~C.}\ \bibnamefont {White}},
  \bibinfo {author} {\bibfnamefont {Y.}~\bibnamefont {Yin}}, \bibinfo {author}
  {\bibfnamefont {J.}~\bibnamefont {Zhao}}, \bibinfo {author} {\bibfnamefont
  {A.~N.}\ \bibnamefont {Cleland}}, \bibinfo {author} {\bibfnamefont {J.~M.}\
  \bibnamefont {Martinis}}, \ and\ \bibinfo {author} {\bibfnamefont {J.~J.~A.}\
  \bibnamefont {Baselmans}},\ }\href {\doibase 10.1063/1.3638063} {\bibfield
  {journal} {\bibinfo  {journal} {Applied Physics Letters}\ }\textbf {\bibinfo
  {volume} {99}},\ \bibinfo {pages} {113507} (\bibinfo {year} {2011})},\
  \Eprint {http://arxiv.org/abs/https://doi.org/10.1063/1.3638063}
  {https://doi.org/10.1063/1.3638063} \BibitemShut {NoStop}%
\bibitem [{\citenamefont {Córcoles}\ \emph {et~al.}(2011)\citenamefont
  {Córcoles}, \citenamefont {Chow}, \citenamefont {Gambetta}, \citenamefont
  {Rigetti}, \citenamefont {Rozen}, \citenamefont {Keefe}, \citenamefont
  {Beth~Rothwell}, \citenamefont {Ketchen},\ and\ \citenamefont
  {Steffen}}]{Corcoles_2011}%
  \BibitemOpen
  \bibfield  {author} {\bibinfo {author} {\bibfnamefont {A.~D.}\ \bibnamefont
  {Córcoles}}, \bibinfo {author} {\bibfnamefont {J.~M.}\ \bibnamefont {Chow}},
  \bibinfo {author} {\bibfnamefont {J.~M.}\ \bibnamefont {Gambetta}}, \bibinfo
  {author} {\bibfnamefont {C.}~\bibnamefont {Rigetti}}, \bibinfo {author}
  {\bibfnamefont {J.~R.}\ \bibnamefont {Rozen}}, \bibinfo {author}
  {\bibfnamefont {G.~A.}\ \bibnamefont {Keefe}}, \bibinfo {author}
  {\bibfnamefont {M.}~\bibnamefont {Beth~Rothwell}}, \bibinfo {author}
  {\bibfnamefont {M.~B.}\ \bibnamefont {Ketchen}}, \ and\ \bibinfo {author}
  {\bibfnamefont {M.}~\bibnamefont {Steffen}},\ }\href {\doibase
  10.1063/1.3658630} {\bibfield  {journal} {\bibinfo  {journal} {Applied
  Physics Letters}\ }\textbf {\bibinfo {volume} {99}},\ \bibinfo {pages}
  {181906} (\bibinfo {year} {2011})},\ \Eprint
  {http://arxiv.org/abs/https://doi.org/10.1063/1.3658630}
  {https://doi.org/10.1063/1.3658630} \BibitemShut {NoStop}%
\bibitem [{\citenamefont {Macklin}\ \emph {et~al.}(2015)\citenamefont
  {Macklin}, \citenamefont {OBrien}, \citenamefont {Hover}, \citenamefont
  {Schwartz}, \citenamefont {Bolkhovsky}, \citenamefont {Zhang}, \citenamefont
  {Oliver},\ and\ \citenamefont {Siddiqi}}]{Macklin_2015}%
  \BibitemOpen
  \bibfield  {author} {\bibinfo {author} {\bibfnamefont {C.}~\bibnamefont
  {Macklin}}, \bibinfo {author} {\bibfnamefont {K.}~\bibnamefont {OBrien}},
  \bibinfo {author} {\bibfnamefont {D.}~\bibnamefont {Hover}}, \bibinfo
  {author} {\bibfnamefont {M.~E.}\ \bibnamefont {Schwartz}}, \bibinfo {author}
  {\bibfnamefont {V.}~\bibnamefont {Bolkhovsky}}, \bibinfo {author}
  {\bibfnamefont {X.}~\bibnamefont {Zhang}}, \bibinfo {author} {\bibfnamefont
  {W.~D.}\ \bibnamefont {Oliver}}, \ and\ \bibinfo {author} {\bibfnamefont
  {I.}~\bibnamefont {Siddiqi}},\ }\href {\doibase 10.1126/science.aaa8525}
  {\bibfield  {journal} {\bibinfo  {journal} {Science}\ }\textbf {\bibinfo
  {volume} {350}},\ \bibinfo {pages} {307} (\bibinfo {year}
  {2015})}\BibitemShut {NoStop}%
\bibitem [{\citenamefont {Tuckerman}\ \emph {et~al.}(2016)\citenamefont
  {Tuckerman}, \citenamefont {Hamilton}, \citenamefont {Reilly}, \citenamefont
  {Bai}, \citenamefont {Hernandez}, \citenamefont {Hornibrook}, \citenamefont
  {Sellers},\ and\ \citenamefont {Ellis}}]{Tuckerman_2016}%
  \BibitemOpen
  \bibfield  {author} {\bibinfo {author} {\bibfnamefont {D.~B.}\ \bibnamefont
  {Tuckerman}}, \bibinfo {author} {\bibfnamefont {M.~C.}\ \bibnamefont
  {Hamilton}}, \bibinfo {author} {\bibfnamefont {D.~J.}\ \bibnamefont
  {Reilly}}, \bibinfo {author} {\bibfnamefont {R.}~\bibnamefont {Bai}},
  \bibinfo {author} {\bibfnamefont {G.~A.}\ \bibnamefont {Hernandez}}, \bibinfo
  {author} {\bibfnamefont {J.~M.}\ \bibnamefont {Hornibrook}}, \bibinfo
  {author} {\bibfnamefont {J.~A.}\ \bibnamefont {Sellers}}, \ and\ \bibinfo
  {author} {\bibfnamefont {C.~D.}\ \bibnamefont {Ellis}},\ }\href {\doibase
  10.1088/0953-2048/29/8/084007} {\bibfield  {journal} {\bibinfo  {journal}
  {Superconductor Science and Technology}\ }\textbf {\bibinfo {volume} {29}},\
  \bibinfo {pages} {084007} (\bibinfo {year} {2016})}\BibitemShut {NoStop}%
\bibitem [{\citenamefont {{Walter}}\ \emph {et~al.}(2018)\citenamefont
  {{Walter}}, \citenamefont {{Bockstiegel}}, \citenamefont {{Mazin}},\ and\
  \citenamefont {{Daal}}}]{Walter_2018}%
  \BibitemOpen
  \bibfield  {author} {\bibinfo {author} {\bibfnamefont {A.~B.}\ \bibnamefont
  {{Walter}}}, \bibinfo {author} {\bibfnamefont {C.}~\bibnamefont
  {{Bockstiegel}}}, \bibinfo {author} {\bibfnamefont {B.~A.}\ \bibnamefont
  {{Mazin}}}, \ and\ \bibinfo {author} {\bibfnamefont {M.}~\bibnamefont
  {{Daal}}},\ }\href {\doibase 10.1109/TASC.2017.2773836} {\bibfield  {journal}
  {\bibinfo  {journal} {IEEE Transactions on Applied Superconductivity}\
  }\textbf {\bibinfo {volume} {28}},\ \bibinfo {pages} {1} (\bibinfo {year}
  {2018})}\BibitemShut {NoStop}%
\bibitem [{\citenamefont {Das}\ \emph {et~al.}(2018)\citenamefont {Das},
  \citenamefont {Yoder}, \citenamefont {Rosenberg}, \citenamefont {Kim},
  \citenamefont {Yost}, \citenamefont {Mallek}, \citenamefont {Hover},
  \citenamefont {Bolkhovsky}, \citenamefont {Kerman},\ and\ \citenamefont
  {Oliver}}]{Das_2018}%
  \BibitemOpen
  \bibfield  {author} {\bibinfo {author} {\bibfnamefont {R.~N.}\ \bibnamefont
  {Das}}, \bibinfo {author} {\bibfnamefont {J.~L.}\ \bibnamefont {Yoder}},
  \bibinfo {author} {\bibfnamefont {D.}~\bibnamefont {Rosenberg}}, \bibinfo
  {author} {\bibfnamefont {D.}~\bibnamefont {Kim}}, \bibinfo {author}
  {\bibfnamefont {D.-R.}\ \bibnamefont {Yost}}, \bibinfo {author}
  {\bibfnamefont {J.}~\bibnamefont {Mallek}}, \bibinfo {author} {\bibfnamefont
  {D.~J.}\ \bibnamefont {Hover}}, \bibinfo {author} {\bibfnamefont
  {V.}~\bibnamefont {Bolkhovsky}}, \bibinfo {author} {\bibfnamefont {A.~J.}\
  \bibnamefont {Kerman}}, \ and\ \bibinfo {author} {\bibfnamefont
  {W.}~\bibnamefont {Oliver}},\ }\href@noop {} {\bibfield  {journal} {\bibinfo
  {journal} {2018 IEEE 68th Electronic Components and Technology Conference
  (ECTC)}\ ,\ \bibinfo {pages} {504}} (\bibinfo {year} {2018})}\BibitemShut
  {NoStop}%
\bibitem [{\citenamefont {Kamal}\ \emph {et~al.}(2014)\citenamefont {Kamal},
  \citenamefont {Roy}, \citenamefont {Clarke},\ and\ \citenamefont
  {Devoret}}]{Kamal_2014}%
  \BibitemOpen
  \bibfield  {author} {\bibinfo {author} {\bibfnamefont {A.}~\bibnamefont
  {Kamal}}, \bibinfo {author} {\bibfnamefont {A.}~\bibnamefont {Roy}}, \bibinfo
  {author} {\bibfnamefont {J.}~\bibnamefont {Clarke}}, \ and\ \bibinfo {author}
  {\bibfnamefont {M.~H.}\ \bibnamefont {Devoret}},\ }\href {\doibase
  10.1103/PhysRevLett.113.247003} {\bibfield  {journal} {\bibinfo  {journal}
  {Phys. Rev. Lett.}\ }\textbf {\bibinfo {volume} {113}},\ \bibinfo {pages}
  {247003} (\bibinfo {year} {2014})}\BibitemShut {NoStop}%
\bibitem [{\citenamefont {Chapman}\ \emph {et~al.}(2018)\citenamefont
  {Chapman}, \citenamefont {Rosenthal},\ and\ \citenamefont
  {Lehnert}}]{Chapman_2018}%
  \BibitemOpen
  \bibfield  {author} {\bibinfo {author} {\bibfnamefont {B.~J.}\ \bibnamefont
  {Chapman}}, \bibinfo {author} {\bibfnamefont {E.~I.}\ \bibnamefont
  {Rosenthal}}, \ and\ \bibinfo {author} {\bibfnamefont {K.~W.}\ \bibnamefont
  {Lehnert}},\ }\href@noop {} {\enquote {\bibinfo {title} {Design of an on-chip
  superconducting microwave circulator with octave bandwidth},}\ } (\bibinfo
  {year} {2018}),\ \Eprint {http://arxiv.org/abs/1809.08747} {arXiv:1809.08747
  [quant-ph]} \BibitemShut {NoStop}%
\bibitem [{\citenamefont {of~Sciences~Engineering}\ and\ \citenamefont
  {Medicine}(2018)}]{NAP25196}%
  \BibitemOpen
  \bibfield  {author} {\bibinfo {author} {\bibfnamefont {N.~A.}\ \bibnamefont
  {of~Sciences~Engineering}}\ and\ \bibinfo {author} {\bibnamefont
  {Medicine}},\ }\href {\doibase 10.17226/25196} {\emph {\bibinfo {title}
  {Quantum Computing: Progress and Prospects}}},\ edited by\ \bibinfo {editor}
  {\bibfnamefont {E.}~\bibnamefont {Grumbling}}\ and\ \bibinfo {editor}
  {\bibfnamefont {M.}~\bibnamefont {Horowitz}}\ (\bibinfo  {publisher} {The
  National Academies Press},\ \bibinfo {year} {2018})\BibitemShut {NoStop}%
\bibitem [{\citenamefont {Bardin}\ \emph {et~al.}(2019)\citenamefont {Bardin},
  \citenamefont {Jeffrey}, \citenamefont {Lucero}, \citenamefont {Huang},
  \citenamefont {Naaman}, \citenamefont {Barends}, \citenamefont {White},
  \citenamefont {Giustina}, \citenamefont {Sank}, \citenamefont {Roushan},
  \citenamefont {Arya}, \citenamefont {Chiaro}, \citenamefont {Kelly},
  \citenamefont {Chen}, \citenamefont {Burkett}, \citenamefont {Chen},
  \citenamefont {Dunsworth}, \citenamefont {Fowler}, \citenamefont {Foxen},
  \citenamefont {Gidney}, \citenamefont {Graff}, \citenamefont {Klimov},
  \citenamefont {Mutus}, \citenamefont {McEwen}, \citenamefont {Megrant},
  \citenamefont {Neeley}, \citenamefont {Neill}, \citenamefont {Quintana},
  \citenamefont {Vainsencher}, \citenamefont {Neven},\ and\ \citenamefont
  {Martinis}}]{Bardin_2019}%
  \BibitemOpen
  \bibfield  {author} {\bibinfo {author} {\bibfnamefont {J.}~\bibnamefont
  {Bardin}}, \bibinfo {author} {\bibfnamefont {E.}~\bibnamefont {Jeffrey}},
  \bibinfo {author} {\bibfnamefont {E.}~\bibnamefont {Lucero}}, \bibinfo
  {author} {\bibfnamefont {T.}~\bibnamefont {Huang}}, \bibinfo {author}
  {\bibfnamefont {O.}~\bibnamefont {Naaman}}, \bibinfo {author} {\bibfnamefont
  {R.}~\bibnamefont {Barends}}, \bibinfo {author} {\bibfnamefont
  {T.}~\bibnamefont {White}}, \bibinfo {author} {\bibfnamefont
  {M.}~\bibnamefont {Giustina}}, \bibinfo {author} {\bibfnamefont
  {D.}~\bibnamefont {Sank}}, \bibinfo {author} {\bibfnamefont {P.}~\bibnamefont
  {Roushan}}, \bibinfo {author} {\bibfnamefont {K.}~\bibnamefont {Arya}},
  \bibinfo {author} {\bibfnamefont {B.}~\bibnamefont {Chiaro}}, \bibinfo
  {author} {\bibfnamefont {J.}~\bibnamefont {Kelly}}, \bibinfo {author}
  {\bibfnamefont {J.}~\bibnamefont {Chen}}, \bibinfo {author} {\bibfnamefont
  {B.}~\bibnamefont {Burkett}}, \bibinfo {author} {\bibfnamefont
  {Y.}~\bibnamefont {Chen}}, \bibinfo {author} {\bibfnamefont {A.}~\bibnamefont
  {Dunsworth}}, \bibinfo {author} {\bibfnamefont {A.}~\bibnamefont {Fowler}},
  \bibinfo {author} {\bibfnamefont {B.}~\bibnamefont {Foxen}}, \bibinfo
  {author} {\bibfnamefont {C.~M.}\ \bibnamefont {Gidney}}, \bibinfo {author}
  {\bibfnamefont {R.}~\bibnamefont {Graff}}, \bibinfo {author} {\bibfnamefont
  {P.}~\bibnamefont {Klimov}}, \bibinfo {author} {\bibfnamefont
  {J.}~\bibnamefont {Mutus}}, \bibinfo {author} {\bibfnamefont
  {M.}~\bibnamefont {McEwen}}, \bibinfo {author} {\bibfnamefont
  {A.}~\bibnamefont {Megrant}}, \bibinfo {author} {\bibfnamefont
  {M.}~\bibnamefont {Neeley}}, \bibinfo {author} {\bibfnamefont
  {C.}~\bibnamefont {Neill}}, \bibinfo {author} {\bibfnamefont
  {C.}~\bibnamefont {Quintana}}, \bibinfo {author} {\bibfnamefont
  {A.}~\bibnamefont {Vainsencher}}, \bibinfo {author} {\bibfnamefont
  {H.}~\bibnamefont {Neven}}, \ and\ \bibinfo {author} {\bibfnamefont
  {J.}~\bibnamefont {Martinis}},\ }in\ \href@noop {} {\emph {\bibinfo
  {booktitle} {Proceedings of the 2019 International Solid State Circuits
  Conference}}}\ (\bibinfo {year} {2019})\ pp.\ \bibinfo {pages}
  {456--458}\BibitemShut {NoStop}%
\bibitem [{\citenamefont {Fedorov}\ \emph {et~al.}(2014)\citenamefont
  {Fedorov}, \citenamefont {Shcherbakova}, \citenamefont {Wolf}, \citenamefont
  {Beckmann},\ and\ \citenamefont {Ustinov}}]{Kirill_2014}%
  \BibitemOpen
  \bibfield  {author} {\bibinfo {author} {\bibfnamefont {K.~G.}\ \bibnamefont
  {Fedorov}}, \bibinfo {author} {\bibfnamefont {A.~V.}\ \bibnamefont
  {Shcherbakova}}, \bibinfo {author} {\bibfnamefont {M.~J.}\ \bibnamefont
  {Wolf}}, \bibinfo {author} {\bibfnamefont {D.}~\bibnamefont {Beckmann}}, \
  and\ \bibinfo {author} {\bibfnamefont {A.~V.}\ \bibnamefont {Ustinov}},\
  }\href {\doibase 10.1103/PhysRevLett.112.160502} {\bibfield  {journal}
  {\bibinfo  {journal} {Phys. Rev. Lett.}\ }\textbf {\bibinfo {volume} {112}},\
  \bibinfo {pages} {160502} (\bibinfo {year} {2014})}\BibitemShut {NoStop}%
\bibitem [{\citenamefont {Leonard}\ \emph {et~al.}(2019)\citenamefont
  {Leonard}, \citenamefont {Beck}, \citenamefont {Nelson}, \citenamefont
  {Christensen}, \citenamefont {Thorbeck}, \citenamefont {Howington},
  \citenamefont {Opremcak}, \citenamefont {Pechenezhskiy}, \citenamefont
  {Dodge}, \citenamefont {Dupuis}, \citenamefont {Hutchings}, \citenamefont
  {Ku}, \citenamefont {Schlenker}, \citenamefont {Suttle}, \citenamefont
  {Wilen}, \citenamefont {Zhu}, \citenamefont {Vavilov}, \citenamefont
  {Plourde},\ and\ \citenamefont {McDermott}}]{Leonard_2019}%
  \BibitemOpen
  \bibfield  {author} {\bibinfo {author} {\bibfnamefont {E.}~\bibnamefont
  {Leonard}}, \bibinfo {author} {\bibfnamefont {M.~A.}\ \bibnamefont {Beck}},
  \bibinfo {author} {\bibfnamefont {J.}~\bibnamefont {Nelson}}, \bibinfo
  {author} {\bibfnamefont {B.}~\bibnamefont {Christensen}}, \bibinfo {author}
  {\bibfnamefont {T.}~\bibnamefont {Thorbeck}}, \bibinfo {author}
  {\bibfnamefont {C.}~\bibnamefont {Howington}}, \bibinfo {author}
  {\bibfnamefont {A.}~\bibnamefont {Opremcak}}, \bibinfo {author}
  {\bibfnamefont {I.}~\bibnamefont {Pechenezhskiy}}, \bibinfo {author}
  {\bibfnamefont {K.}~\bibnamefont {Dodge}}, \bibinfo {author} {\bibfnamefont
  {N.}~\bibnamefont {Dupuis}}, \bibinfo {author} {\bibfnamefont
  {M.}~\bibnamefont {Hutchings}}, \bibinfo {author} {\bibfnamefont
  {J.}~\bibnamefont {Ku}}, \bibinfo {author} {\bibfnamefont {F.}~\bibnamefont
  {Schlenker}}, \bibinfo {author} {\bibfnamefont {J.}~\bibnamefont {Suttle}},
  \bibinfo {author} {\bibfnamefont {C.}~\bibnamefont {Wilen}}, \bibinfo
  {author} {\bibfnamefont {S.}~\bibnamefont {Zhu}}, \bibinfo {author}
  {\bibfnamefont {M.}~\bibnamefont {Vavilov}}, \bibinfo {author} {\bibfnamefont
  {B.}~\bibnamefont {Plourde}}, \ and\ \bibinfo {author} {\bibfnamefont
  {R.}~\bibnamefont {McDermott}},\ }\href {\doibase
  10.1103/PhysRevApplied.11.014009} {\bibfield  {journal} {\bibinfo  {journal}
  {Phys. Rev. Applied}\ }\textbf {\bibinfo {volume} {11}},\ \bibinfo {pages}
  {014009} (\bibinfo {year} {2019})}\BibitemShut {NoStop}%
\bibitem [{\citenamefont {McDermott}\ and\ \citenamefont
  {Vavilov}(2014)}]{McDermott_2014}%
  \BibitemOpen
  \bibfield  {author} {\bibinfo {author} {\bibfnamefont {R.}~\bibnamefont
  {McDermott}}\ and\ \bibinfo {author} {\bibfnamefont {M.~G.}\ \bibnamefont
  {Vavilov}},\ }\href {\doibase 10.1103/PhysRevApplied.2.014007} {\bibfield
  {journal} {\bibinfo  {journal} {Phys. Rev. Applied}\ }\textbf {\bibinfo
  {volume} {2}},\ \bibinfo {pages} {014007} (\bibinfo {year}
  {2014})}\BibitemShut {NoStop}%
\bibitem [{\citenamefont {Liebermann}\ and\ \citenamefont
  {Wilhelm}(2016)}]{Liebermann_2016}%
  \BibitemOpen
  \bibfield  {author} {\bibinfo {author} {\bibfnamefont {P.~J.}\ \bibnamefont
  {Liebermann}}\ and\ \bibinfo {author} {\bibfnamefont {F.~K.}\ \bibnamefont
  {Wilhelm}},\ }\href {\doibase 10.1103/PhysRevApplied.6.024022} {\bibfield
  {journal} {\bibinfo  {journal} {Phys. Rev. Applied}\ }\textbf {\bibinfo
  {volume} {6}},\ \bibinfo {pages} {024022} (\bibinfo {year}
  {2016})}\BibitemShut {NoStop}%
\bibitem [{\citenamefont {Kimble}(2018)}]{Kimble_2018}%
  \BibitemOpen
  \bibfield  {author} {\bibinfo {author} {\bibfnamefont {H.~J.}\ \bibnamefont
  {Kimble}},\ }\href {\doibase 10.1038/nature07127} {\bibfield  {journal}
  {\bibinfo  {journal} {Nature}\ }\textbf {\bibinfo {volume} {453}},\ \bibinfo
  {pages} {1023} (\bibinfo {year} {2018})}\BibitemShut {NoStop}%
\bibitem [{\citenamefont {Han}\ \emph {et~al.}(2018)\citenamefont {Han},
  \citenamefont {Vogt}, \citenamefont {Gross}, \citenamefont {Jaksch},
  \citenamefont {Kiffner},\ and\ \citenamefont {Li}}]{Han_2018}%
  \BibitemOpen
  \bibfield  {author} {\bibinfo {author} {\bibfnamefont {J.}~\bibnamefont
  {Han}}, \bibinfo {author} {\bibfnamefont {T.}~\bibnamefont {Vogt}}, \bibinfo
  {author} {\bibfnamefont {C.}~\bibnamefont {Gross}}, \bibinfo {author}
  {\bibfnamefont {D.}~\bibnamefont {Jaksch}}, \bibinfo {author} {\bibfnamefont
  {M.}~\bibnamefont {Kiffner}}, \ and\ \bibinfo {author} {\bibfnamefont
  {W.}~\bibnamefont {Li}},\ }\href {\doibase 10.1103/PhysRevLett.120.093201}
  {\bibfield  {journal} {\bibinfo  {journal} {Phys. Rev. Lett.}\ }\textbf
  {\bibinfo {volume} {120}},\ \bibinfo {pages} {093201} (\bibinfo {year}
  {2018})}\BibitemShut {NoStop}%
\bibitem [{\citenamefont {Andrews}\ \emph {et~al.}(2014)\citenamefont
  {Andrews}, \citenamefont {Peterson}, \citenamefont {Purdy}, \citenamefont
  {Cicak}, \citenamefont {Simmonds}, \citenamefont {Regal},\ and\ \citenamefont
  {Lehnert}}]{Andrews_2014}%
  \BibitemOpen
  \bibfield  {author} {\bibinfo {author} {\bibfnamefont {R.~W.}\ \bibnamefont
  {Andrews}}, \bibinfo {author} {\bibfnamefont {R.~W.}\ \bibnamefont
  {Peterson}}, \bibinfo {author} {\bibfnamefont {T.~P.}\ \bibnamefont {Purdy}},
  \bibinfo {author} {\bibfnamefont {K.}~\bibnamefont {Cicak}}, \bibinfo
  {author} {\bibfnamefont {R.~W.}\ \bibnamefont {Simmonds}}, \bibinfo {author}
  {\bibfnamefont {C.~A.}\ \bibnamefont {Regal}}, \ and\ \bibinfo {author}
  {\bibfnamefont {K.~W.}\ \bibnamefont {Lehnert}},\ }\href@noop {} {\bibfield
  {journal} {\bibinfo  {journal} {Nature Physics}\ }\textbf {\bibinfo {volume}
  {10}},\ \bibinfo {pages} {321} (\bibinfo {year} {2014})}\BibitemShut
  {NoStop}%
\end{thebibliography}%

\end{document}